\documentclass[aps,preprint,showpacs]{revtex4}
\usepackage{epsfig,amssymb,bm}

\begin{document}

\title{
Effective Lagrangian approach to the  \\
$\bm{\omega}$ photoproduction near threshold }


 \author{Alexander I. Titov}
 \altaffiliation
 {Bogoliubov Laboratory of Theoretical Physics, JINR,
  Dubna 141980, Russia}
 \affiliation
 {Advanced Science Research Center, JAERI, Tokai, Ibaraki, 319-1195
  Japan}
 \email{atitov@thsun1.jinr.ru}
 \author{T.-S. H. Lee}
 \affiliation{Physics Division, Argonne National Laboratory, Argonne,
 Illinois 60439, USA}
 \email{lee@theory.phy.anl.gov}



\begin{abstract}
We apply the effective Lagrangian approach to investigate the
role of the nucleon resonances in $\omega$ meson photoproduction
at energies near the threshold. The non-resonant amplitudes are
taken from the previous investigations at higher energies and
consist of the pseudoscalar meson exchange and the nucleon Born
terms. The resonant amplitudes are calculated from effective
Lagrangians with the $N^* \rightarrow \gamma N$ and $N^*
\rightarrow \omega N$ coupling constants fixed by the empirical
helicity amplitudes and
 the vector meson dominance model.
The contributions from the nucleon resonances are found to
be significant in changing the differential cross
sections in a wide interval of $t$
and various spin observables.
In particular, we suggest that a crucial test of our predictions
can be made by measuring single and double spin asymmetries.
\end{abstract}


 \maketitle

\section{Introduction}
 The study of the decays of a nucleon resonance ($N^*$) into
 a nucleon ($N$) and a vector meson ($V=\omega,\rho,\phi$)
 is closely related to several aspects
 of intermediate and
 high  energy physics, ranging
 from resolving the so-called  "missing" resonance
 problem~\cite{IK77-80} to estimating the medium modifications
 on the vector meson
 properties~\cite{FP97,Brown_Rho,Hatsuda_Lee,Rapp_Wambach}.
 Electromagnetic production of vector mesons
 is one of the most promising  reactions to
 determine the $N^*\to N V$ couplings
 experimentally, e.g., at  Thomas Jefferson National Accelerator Facility,
 ELSA-SAPHIR of Bonn, GRAAL of Grenoble, and LEPS of SPring-8.
 The $\omega$ meson photoproduction has some advantages because the
 non-resonant contribution to its total amplitude
 is much better understood
 as compared with the other vector mesons\cite{FS96}.

 The role of the nucleon resonances  in $\omega$ meson
 photoproduction at relatively large energy was studied  by Zhao
 {\em et al.\/} \cite{ZLB98,Zhao99,ZDGS99}, using the $\mbox{SU}(6)
 \times \mbox{O}(3)$ constituent quark model with an effective
 quark-meson interaction. By fitting the model parameters to the
 existing data, they found that the single polarization observables
 are sensitive to the nucleon resonances and the dominant
 contribution at $E_\gamma\simeq 1.7$ GeV comes from the "missing"
 $F_{15}(2000)$ resonance. A recent analysis\cite{OTL01} of
 $\omega$ photoproduction in the same energy region makes use of
 the constituent quark model~\cite{Caps92,CR94},
 which accounts for the configuration mixing due to the
 residual quark-quark interactions~\cite{GI85-CI86}.
 In~\cite{OTL01} the dominant contributions are found to be from $N\frac32^+ (1910)$,
 a missing resonance, and $N\frac32^- (1960)$, which is identified as
 the $D_{13}(2080)$ of the Particle Data Group(PDG)~\cite{PDG98}.
 As a result, the predicted single polarizations are rather
 different from those of~\cite{ZLB98}. Hopefully, the data from near
 future will clarify the situation.

It would be important to extend these studies to the low energy
region, $E_\gamma\sim 1.1-1.25$ GeV,
close to the omega production threshold,
where the well established three- and four-star nucleon resonances
are expected to be important. Unfortunately
neither the approach of Ref.~\cite{ZLB98}, nor of Ref.~\cite{OTL01}
 could be applied for this investigation.
The first approach\cite{ZLB98}  does not
include the configuration mixing. Therefore, the predicted
low-lying $N^*$ states, belonging to the
$[{}^4{\bf 8,70}]$ representation, can not contribute
to the resonance photo excitation on the proton target because
of the Moorhouse selection rule.
That is, the contributions from  $S_{11}(1650),\, D_{15}(1675)$,
and $D_{13}(1700)$ resonances are strictly suppressed~\cite{Close}.
This contradicts with the experimental data~\cite{PDG98}, which, for
 example, shows that  photo excitation  helicity
amplitude $A^{\frac12}_{p}$ for $S_{11}(1650)$ resonance is
 finite and large.
In general, one finds that the SU(6)$\times$O(3)
quark model without configuration mixing
forbids or suppresses most of the low mass resonances.

While the second approach\cite{OTL01} makes use of the information
from the constituent quark model with configuration mixing, the
$N^*\rightarrow \omega N$ transition amplitudes for the low-lying
$N^*$'s with mass less than 1.72 GeV are not  available in
Ref.\cite{CR94}. To provide such information, some extensions of
the $^3P_0$ model employed in Ref.\cite{CR94} for predicting
$N^*\rightarrow \omega N$ are necessary.

As a possible solution of the problems mentioned above, we will
follow the previous studies of $\pi$~\cite{DMW91,SL96}
and $\eta$-photoproduction~\cite{BMZ95} and
 apply the  effective Lagrangian approach to also describe
the $N^*$ excitation.
Our approach is similar to the recent work by Riska and
Brown\cite{RB00}, but with several significant differences.
They fix the $N^* N \omega(\rho)$ effective Lagrangians
for the low mass resonances with $M_{N^*}\leq 1720$ MeV by
using the matrix elements derived from the SU(6)$\times$O(3)
constituent quark model with no configuration mixing.
For the reasons mentioned above, an extension
 of their approach to also define the effective Lagrangian
for  $N^*\rightarrow \gamma N$ coupling will
be unrealistic because of the restriction of the
SU(6)$\times$O(3) quark model.
 Thus their model is not applicable
to a consistent  investigation using effective Lagrangian approach.
In this work, we will start with the
empirical $N^*\rightarrow \gamma N$ amplitudes and then
predict the needed $N^*\rightarrow \omega N$ coupling constants by using
the vector meson dominance assumption. This is similar to a recent work
by Post and Mosel\cite{PM00}.
But we will use a fully covariant formulation and
explore in more detail the consequesncies of this approach.

To proceed, we need to also consider non-resonant mechanisms.
It is fairly well established
that the $\omega$ meson  photoproduction
is dominated by diffractive processes at high energies
and by the one-pion exchange and the standard
nucleon Born terms at low energies.
The diffractive part is associated with  the Pomeron exchange.
In the near threshold energy the Pomeron exchange contribution
is negligible and can be safely neglected in this work.

The calculation of the one-pion exchange amplitude(Fig.1a) and the
direct and crossed nucleon terms(Figs. 1b and 1c) has been
recently revived in Refs.\cite{FS96,OTL01}. In this work we use
the parameters determined in Ref.\cite{OTL01} to evaluate these
non-resonant amplitudes. Our focus is to construct effective
Lagrangians for calculating the  resonant amplitude shown in
Fig.~1(d,e) and explore their consequences in determining various
spin observables.

This paper is organized as follows.
In Section II we define the kinematics and the non-resonant
amplitudes.  Formula for calculating various spin
observables will also be introduced there.
 The effective Lagrangians for calculating the amplitudes
due to the excitations of  nucleon resonances will
be presented in Section III. In Section IV we present results and
predictions for future experimental tests.  The summary is given
in Section V.

\section{Kinematics, Non-resonant amplitude and Observables}

The scattering amplitude $T$ of $\gamma p \rightarrow \omega p$
reaction is related to the $S$-matrix by
\begin{equation}
S^{}_{fi} = \delta^{}_{fi}
- i(2\pi)^4 \delta^4(k + p - q - p') T^{}_{fi},
\label{S:conv}
\end{equation}
where $k$, $q$, $p$ and $p'$ denote the four-momenta of the
incoming photon, outgoing $\omega$, initial nucleon, and final
nucleon respectively. The standard Mandelstam variables are
defined as  $t = (p - p')^2 = (q-k)^2$, $s \equiv W^2 = (p+k)^2$,
and the $\omega$ production angle $\theta$ by $\cos\theta \equiv
{\bf k} \cdot {\bf q} / |{\bf k}| |{\bf q}|$. The scattering
amplitude is written as
\begin{equation}
T^{}_{fi} = \frac{I_{fi}} {(2\pi)^6\,
\sqrt{ 2E^{}_\omega({\bf q})\,2|{\bf k}|
2E_{N}({\bf p}) 2E_{N}({\bf p}')} },
\label{T:conv}
\end{equation}
where $E^{}_\alpha({\bf p}) = \sqrt{M^2_\alpha + {\bf p}^2}$ with
$M^{}_\alpha$ denoting the mass of the particle $\alpha$.
The invariant amplitude has two components
\begin{eqnarray}
I^{}_{fi} = I^{BG}_{fi} + I^{N^*}_{fi},
\end{eqnarray}
where the  resonance excitation term $I^{N^*}_{fi}$
will be  derived in Sect.~III.
The non-resonant (background) amplitude is
\begin{eqnarray}
I^{BG}_{fi} = I^{PS}_{fi} + I^{N}_{fi},
\end{eqnarray}
with  $I^{PS}_{fi}$, and $I^{N}_{fi}$ denoting the amplitudes
due to  the pseudoscalar ($\pi,\eta$)  meson exchange(Fig.1a), and
the direct and crossed nucleon terms(Figs.1b and 1c), respectively.
 We calculate  $I^{PS}_{fi}$ by using the
following effective Lagrangians
\begin{eqnarray}
&&{\cal L}_{\omega\gamma \varphi} =\frac{e}{M_\omega} g_{\omega\gamma \varphi}
\epsilon^{\mu\nu\alpha\beta}
\partial_\mu \omega_\nu \partial_\alpha A_\beta\, \varphi,\qquad
\label{pivgamma}\\
&&{\cal L}_{PS} =-i
{g_{\pi^0 NN}}
\bar N \gamma_5\,\tau_3\, N \,\pi^0
 -i
{g_{\eta NN}} \bar N \gamma_5\, N\,\eta,
\label{PS}
\end{eqnarray}
where  $\varphi=\pi^0,\eta$, and  $A_\beta$ is the photon field.
We use $g_{\pi NN}/4\pi=14$ and  $g_{\eta NN}/4\pi=0.99$
for the $\pi NN$ and $\eta NN$ couplings,
respectively. The $\eta NN$ coupling
constant has some uncertainties, but our choice is within the
range reported in literatures.
The $\omega\gamma\phi$ coupling constants  can be
estimated from the decay widths of $\omega\to\gamma\pi$
and  $\omega\to\gamma\eta$~\cite{PDG98}. This leads to
$g_{\omega\gamma\pi}=1.823$ and $g_{\omega\gamma\eta}=0.416$.
The $\varphi NN$ and $\omega\gamma\varphi$
vertices are regularized  by the following form factors
\begin{eqnarray}
F_{\varphi NN}(t)
=\frac{\Lambda_\varphi^2 -M^2_\varphi}{\Lambda_\varphi^2 -t},\qquad
F_{\omega\gamma\varphi}(t)
=\frac{{\Lambda_{\omega\gamma\varphi}}^2-M_\varphi^2}
{{\Lambda_{\omega\gamma\varphi}}^2-t}.
\end{eqnarray}

We evaluate the direct and crossed nucleon terms by using the
following interaction Lagrangians
\begin{eqnarray}
{\cal L}_{\gamma NN} & = &
- e \left( \bar N \gamma_\mu\,\frac{1+\tau_3}{2}  N {A}^\mu
-\frac{\kappa_N}{2M_N} \bar N \sigma^{\mu\nu}\,N
\partial_\nu {A}_\mu \right),
\label{L_gammaNN}\\
{\cal L}_{\omega NN} & = &
- g_{\omega NN} \left( \bar N \gamma_\mu  N {\omega}^\mu
-\frac{\kappa_\omega}{2M_N} \bar N \sigma^{\mu\nu}\,N
\partial_\nu {\omega}_\mu \right),
\label{L_omegaNN}
\end{eqnarray}
with $\kappa_{p(n)}=1.79(-1.91)$.
For the coupling constants we use
$g_{\omega NN}=10.35$ and $\kappa_\omega=0$, which are
close to the values determined in a study\cite{SL96}
 of pion photoproduction.
We follow Ref.\cite{Hab97} to assume that
the $\gamma NN$ and $\omega NN$ vertices in Figs.1b and 1c
are regularized by the following form factor
\begin{eqnarray}
F_{B}(r^2) =
\frac{\Lambda_B^4}{\Lambda_N^4 + (r^2-M^2_B)^2 },
\label{cutN}
\end{eqnarray}
where $M_B$ and $r$ are the mass and the four-momentum  of the
intermediate baryon state.

The values of cutoff parameters for the
form factors Eqs.(7) and (10) have been determined in a
study\cite{OTL01}
 of $\gamma p \rightarrow \omega p$
at higher energies. Here we use their values with
$\Lambda_\pi=\Lambda^\pi_{\omega\gamma\pi} = 0.6$ GeV/c,
 $\Lambda_{\eta} = 1$ GeV/c, and
$\Lambda_{\omega\gamma\eta}=0.9$ GeV/c
for Eq.(7), and
$\Lambda_B=\Lambda_N=0.5$ GeV for Eq.(10).
They are comparable to the values used in
the literature and are sufficient for the present investigation.
Other forms of form factors, such as those proposed
recently in Ref.\cite{Workman}, will
not be considered in this work.

We calculate all observables in the center of mass system(c.m.s.).
The differential cross section is related to the invariant amplitude
by
\begin{eqnarray}
\frac{d\sigma_{fi}}{dt}=
\frac{1}{64\pi(W^2-M_N^2)^2}\sum_{\lambda_i\lambda_f\lambda_\gamma\lambda_\omega}
|I_{fi}|^2,
\label{cs}
\end{eqnarray}
where $\lambda_i,\lambda_\gamma$ are the helicities of the
incoming nucleon
and photon, and $\lambda_f,\lambda_\omega$ are the helicities of the
outgoing nucleon and $\omega$ meson. In this paper we will also
investigate
the single and double spin observables~\cite{PST96}.
The considered single observables,  vector
beam ($\Sigma_x$), target ($T_y$) and recoil ($P_y$) asymmetries,
are defined by
\begin{eqnarray}
\Sigma_x&=&
\frac{{\rm Tr}\left[I_{fi}\sigma_\gamma^x\,I_{fi}^\dagger\right]}
{{\rm Tr}\left[I_{fi}\,I_{fi}^\dagger\right]}
=\frac{d\sigma_\perp -d\sigma_\parallel}{d\sigma_{\rm tot}} ,
\label{sigma_x}\\
T_y&=&
\frac{{\rm Tr}\left[I_{fi}\sigma_N^y\,I_{fi}^\dagger\right]}
{{\rm Tr}\left[I_{fi}\,I_{fi}^\dagger\right]}
=\frac{d\sigma^{s_y^{N}=\frac12} -d\sigma^{s_y^{N}=-\frac12}}{d\sigma_{\rm
tot}},
\label{T_y}\\
P_y&=&
\frac{{\rm Tr}\left[I_{fi}\,I_{fi}^\dagger\,\sigma_{N'}^x\right]}
{{\rm Tr}\left[I_{fi}\,I_{fi}^\dagger\right]}
=\frac{d\sigma^{s_y^{N'}=\frac12} -d\sigma^{s_y^{N'}=-\frac12}}{d\sigma_{\rm
tot}},
\label{P_y}
\end{eqnarray}
where the
 subscript $\perp\,(\parallel)$ corresponds to a photon linearly
polarized along the ${\bf y}$ (${\bf x}$) axis;
$s_y^N=\frac12\,(-\frac12)$
corresponds to the case that the proton polarization is
parallel (antiparallel) to the ${\bf y}$ axis.
At $\omega-$meson production angle $\theta=0,\pi$, the
amplitudes with the orbital momentum projection $m_l\neq 0$ vanish
and the spin conserving argument results in
\begin{eqnarray}
\Sigma_x=T_y=P_y=0,\qquad\qquad {\rm at}\,\,\theta=0.
\label{S_0}
\end{eqnarray}

For the outgoing $\omega$ meson we consider tensor asymmetry
\begin{eqnarray}
V_{z'z'}=
\frac{{\rm Tr}\left[I_{fi}\,I_{fi}^\dagger\,S_{\omega}^{zz}\right]}
{{\rm Tr}\left[I_{fi}\,I_{fi}^\dagger\right]}
=\frac{d\sigma^{s_{z'}^{\omega}=1}+d\sigma^{s_{z'}^{\omega}=-1}
-2d\sigma^{s_{z'}^{\omega}=0}}
{\sqrt2 d\sigma_{\rm tot}},
\end{eqnarray}
where $S_{\omega}^{zz}$ is the tensor polarization operator for a
spin 1 particle~\cite{BLS80} and $s^\omega_z$ is the $\omega$ meson
spin projection along the quantization axis ${\bf z'}$.
For the pseudoscalar $\pi,\eta$- exchange, which is the dominant
non-resonant amplitude, the tensor polarization may be written in a compact form
\begin{eqnarray}
V_{z'z'}^{PS}=
\frac{3}{2\sqrt{2}}\left[
\left(\frac{v_\omega-\cos\theta}{1-v_\omega\cos\theta}\right)^2
-\frac13 \right],
\label{V_ps}
\end{eqnarray}
where $v_\omega$ is the $\omega$-meson velocity in c.m.s..
This expression shows that at $\theta=0$, $V_{z'z'}^{PS}=1/\sqrt{2}$.
The asymmetry $V_{z'z'}^{PS}$ has the simplest form in the
Gottfried-Jackson frame in which the $\omega$-meson is at rest
and the
${\bf z}$ axis is in the direction of the incoming photon momentum.
In this frame the helicity conserving pseudoscalar exchange amplitude
has a simple form
\begin{eqnarray}
I^{PS}_{\lambda_\omega\lambda_\gamma}
\sim \lambda_\gamma\delta_{\lambda_\omega\lambda_\gamma},
\end{eqnarray}
which  shows that the longitudinally polarized $\omega$ mesons are
forbidden. As result, one get the asymmetry $V_{z'z'}^{PS}$
to be constant at all $\theta$
\begin{eqnarray}
V_{z'z'}^{PS}=
\frac{1}{\sqrt{2}}.
\label{VGJ}
\end{eqnarray}
A deviation from this pseudoscalar meson-exchange value will
measure the contributions from other mechanisms. We will see that
the contribution of other channels  illustrated in Fig.1,
especially at large $\theta$, leads to deviation from this value.

We will also consider  the beam-target double asymmetry, defined as
\begin{equation}
C^{BT}_{zz} =
\frac{d\sigma(\uparrow\downarrow) - d\sigma(\uparrow\uparrow)}
{d\sigma(\uparrow\downarrow) + d\sigma(\uparrow\uparrow)},
\label{C-BT}
\end{equation}
where the arrows represent the helicities of the incoming photon and
the target protons.

\section{Excitation of nucleon resonances}

 For evaluating the resonant amplitude $I^{N^*_L}_{fi}$, we
 consider the isospin $I=1/2$ nucleon resonances
listed by PDG\cite{PDG98}.
However only the resonances with the empirical
helicity amplitudes of $\gamma N \rightarrow N^*$ transitions
given by PDG can  be included in our approach, as
will be clear below.
We thus have
contributions from 12
 resonances: $P_{11}(1440)$,
 $D_{13}(1520)$, $S_{11}(1535)$,
 $S_{11}(1650)$, $D_{15}(1675)$, $F_{15}(1680)$, $D_{13}(1700)$, $P_{11}(1710)$,
 $P_{13}(1720)$, $F_{17}(1990)$, $D_{13}(2080)$, and $G_{17}(2190)$.
Our first task here is to define effective
 Lagrangians for reproducing the empirical
helicity amplitudes for the $\gamma N\rightarrow N^*$
transitions for these resonances.

 For the $N^{*}$ with spin $J=\frac{1}{2}$,
 the effective Lagrangians for
 the $\gamma NN^*$ interactions
 are chosen to be of the form of the usual
 $\gamma NN$ interaction. But only the
 tensor coupling is kept, since the vector coupling
 violates the gauge invariance for the considered
 $M_{N^*}\neq M_N$ cases. This
 "minimal" form of Lagrangian, previously used in the study of
 $\eta$-photoproduction~\cite{BMZ95}, is
 \begin{eqnarray}
 {\cal L}_{\gamma NN^*}^{\frac{1}{2}^{\pm}}
 =
 \,\, \frac{eg_{\gamma NN^*}}{2M_{N^*}} \bar\psi_{N^*} \,\Gamma^{(\pm)}
 \sigma_{\mu\nu} F^{\mu\nu} \psi_{N} +{\rm h.c.},
  \label{R1/2}
 \end{eqnarray}
 where $\psi_{N},\psi_{N^*}$ and $A_\mu$ are the nucleon,
 nucleon resonance, and photon fields, respectively,
 and $F^{\mu\nu}=\partial^\nu A^\mu - \partial^\mu A^\nu$.
 The coupling
 $\Gamma^+ =1(\Gamma^-=\gamma_5$) defines the excitation
 of a positive(negative) parity $N^*$ state.

 For the $N^{*}$ with spin $J={\frac{3}{2}}$,
 we use the expression introduced in
 Refs.~\cite{DMW91,BMZ95,OO75}
 \begin{eqnarray}
 {\cal L}_{\gamma NN^*}^{\frac{3}{2}^\pm}
  =
 i\frac{eg^{}_{\gamma NN^*}}{M_{N^*}}
 \bar\psi_{N^*}^\mu \,O_{\mu\nu}(Z)\gamma_\lambda\Gamma^{(\mp)}
 F^{\lambda\nu}
 {\psi_{N}}\qquad + {\rm h.c.},
 \label{R3/2}
 \end{eqnarray}
 where
 $\psi_\alpha$ is the Rarita-Schwinger spin-$\frac32$ baryon field.
 The off-shell operator $O_{\mu\nu}(Z)$ is
 \begin{eqnarray}
 O_{\mu\nu}(Z)=g_{\mu\nu}+\left[
 \frac12(1+4Z)A+ Z \right]\gamma_\mu\gamma_\nu,
 \label{O_munu}
 \end{eqnarray}
where $A$ is an arbitrary parameter, defining the so-called "point
transformation". Since the the physical amplitudes are independent
of $A$, we follow Ref.\cite{BMZ95} to  choose $A=-1$. When both
$N$ and $N^*$ are on-mass-shell, Eq.~(\ref{R3/2}) is equivalent to
the form commonly used\cite{JS73,NBL90} in the investigation of
pion photoproduction.

 The effective Lagrangians for the resonances with
 $J^P=\frac{5}{2}^{\pm}, \frac{7}{2}^{\pm}$ are constructed by the analogy with the
 previous case
\begin{eqnarray}
 {\cal L}_{\gamma NN^*}^{\frac{5}{2}^\pm}
 &=&
 \frac{eg^{}_{\gamma NN^*}}{M_{N^*}^{2}}
 \bar\psi_{N^*}^{\mu\alpha} \,O_{\mu\nu}(Z)\gamma_\lambda\Gamma^{(\pm)}
 (\partial_\alpha F^{\lambda\nu})
 {\psi_{N}}\qquad + {\rm h.c.},
 \label{R5/2}\\
 {\cal L}_{\gamma NN^*}^{\frac{7}{2}^\pm}
 &=&-i
 \frac{eg^{}_{\gamma NN^*}}{M_{N^*}^{3}}
 \bar\psi_{N^*}^{\mu\alpha\beta} \,O_{\mu\nu}(Z)\gamma_\lambda\Gamma^{(\mp)}
 (\partial_\beta\partial_\alpha F^{\lambda\nu})
 {\psi_{N}}\qquad + {\rm h.c.},
 \label{R7/2}
 \end{eqnarray}
 where  $\psi_{\alpha\beta}$, and $\psi_{\alpha\beta\gamma}$
 are the Rarita-Schwinger spin $\frac52$ and $\frac72$
 field, respectively. The interaction (\ref{R5/2}) is a covariant generalization
 of non-relativistic expression used in Ref.~\cite{FP97}.

 We define the $\omega NN^*$ coupling by using the vector dominance
 model (VDM). It amounts to assuming
 that the electromagnetic and vector meson fields are
 related to each other by
\begin{eqnarray}
A_{s}^\mu= \frac{em_\omega^2}{2\gamma_\omega}\,\omega^\mu,\qquad\qquad
A_{v}^\mu= \frac{em_\rho^2}{2\gamma_\rho}\,\rho^\mu,
\label{vdm0}
\end{eqnarray}
 where $A_{s(v)}$ is the isoscalar (isovector) part of electromagnetic
 field. The coupling strengths
 $\gamma_\omega=8.53$ and $\gamma_\rho=2.52$ are
 fixed by the  electromagnetic
 $\omega\to e^+e^-,\,\rho\to e^+e^-$ decays~\cite{FS96}.
 In the VDM approach the effective
 ${\cal L}_{\omega NN^*}$ Lagrangian has the same form as
 the corresponding ${\cal L}_{\gamma NN^*}$ with substitution
 \begin{eqnarray}
 A_\mu\to \omega_\mu,\qquad\qquad eg_{\gamma N N^*}\to  f_\omega,
 \label{vdm1}
 \end{eqnarray}
 with
 \begin{eqnarray}
 f_\omega=2g_s\gamma_\omega.
 \label{vdm2}
 \end{eqnarray}
 The scalar coupling constant $g_s$ is related to the
strengths of the $N^*$ excitations
on the proton($g_p = g_{\gamma p N^*}$)
 and on the neutron($g_n=g_{\gamma n N^*}$)
 \begin{eqnarray}
 g_s=\frac{g_p + g_n}{2}.
 \label{vdm3}
 \end{eqnarray}

With the Lagrangians for $\gamma N N^*$ and
 $\omega N N^*$ couplings specified above, we  can
calculate the invariant amplitudes for the
$N^*$ excitation mechanisms illustrated in Figs. 1d and 1e.
 The resulting invariant
amplitudes have the following form
\begin{eqnarray}
I^{(N^*)}_{fi}=eg_{\gamma NN^*}f_{\omega NN^*}
\bar u(p'){\cal A}^{\mu\nu}(N^*)u(p)
\varepsilon^{\omega*}_\mu \varepsilon^{\gamma}_\nu,
\label{IN*_fi}
\end{eqnarray}
where the operators $\cal A_{\mu\nu}$ are completely defined by
the effective Lagrangians~(\ref{R1/2}) - (\ref{R5/2})
\begin{eqnarray}
 {\cal A}_{\mu\nu}^{\frac12^\pm}
 &=&-\frac{eg_{\gamma NN^*}f_{\omega NN^*}}{M_{N^*}^2}
  \left[
  \gamma_\mu \not\hskip-0.7mm\!{q}\,{\cal P}^{s\pm}(N^*)
  \gamma_\nu\not\hskip-0.7mm\!{k}\,F_{N^*}(s)
 + \gamma_\nu \not\hskip-0.7mm\!{k}\,{\cal P}^{u\pm}(N^*)
  \gamma_\mu\not\hskip-0.7mm\!{q}\,F_{N^*}(u)
\right],
\label{A1/2}\\
\nonumber\\
{\cal A}_{\mu\nu}^{\frac32^\pm}
&=&\mp
   \frac{eg^{}_{\gamma NN^*f^{}_{\omega NN^*}}}{(M_{N^*}^2}
 \Gamma^{(\mp)}\left[\right.
 \overline{W}^{\alpha}_\mu
 {\cal P}^s_{\alpha\beta}(N^*)
 E^{\beta}_{\nu}
 F_{N^*}(s)\nonumber\\
 &&\qquad\qquad\qquad\qquad\qquad +
 \overline{E}^{\alpha}_\mu
 {\cal P}^u_{\alpha\beta}(N^*)
 W^{\beta}_{\nu}
 F_{N^*}(u)\left.\right]\,
 \Gamma^{(\mp)},
 \label{A3/2}\\
 \nonumber\\
 {\cal A}_{\mu\nu}^{\frac52^\pm}
&=&\pm
   \frac{eg^{}_{\gamma NN^*}f^{}_{\omega NN^*}}{M_{N^*}^4}
 \Gamma^{(\pm)}\left[\right.
 \overline{W}^{\alpha}_\mu
 q^{\alpha'} {\cal P}^s_{\alpha\alpha',\beta\beta'}(N^*)k^{\beta'}
 E^{\beta}_{\nu}
 F_{N^*}(s)
 \nonumber\\
 &&\qquad\qquad\qquad\qquad\qquad +
 \overline{E}^{\alpha}_\mu
 k^{\alpha'}{\cal P}^u_{\alpha\alpha',\beta\beta'}(N^*)q^{\beta'}
 W^{\beta}_{\nu}
 F_{N^*}(u)
 \left.\right]\Gamma^{(\pm)}
 \label{A5/2}\\
 {\cal A}_{\mu\nu}^{\frac72^\pm}
 &=&\mp
   \frac{eg^{}_{\gamma NN^*}f^{}_{\omega NN^*}}{M_{N^*}^6}
 \Gamma^{(\mp)}\left[\right.
 \overline{W}^{\alpha}_\mu
 q^{\alpha'}  q^{\alpha{''}}
  {\cal P}^s_{\alpha\alpha'\alpha'',\beta\beta'\beta''}(N^*)
  k^{\beta'} k^{\beta''}
 E^{\beta}_{\nu}
 F_{N^*}(s)
 \nonumber\\
 &&\qquad\qquad\qquad\qquad\qquad +
 \overline{E}^{\alpha}_\mu
 k^{\alpha'}k^{\alpha''}
 {\cal P}^u_{\alpha\alpha'\alpha'',\beta\beta'\beta''}(N^*)
 q^{\beta'}q^{\beta''}
 W^{\beta}_{\nu}
 F_{N^*}(u)
 \left.\right]\Gamma^{(\mp)}
 \label{A7/2}
 \end{eqnarray}
In the above equations,
we have introduced
\begin{eqnarray}
\overline{W}^{\alpha}_\mu
&=&
 (\not\hskip-0.7mm\!{q}\,g_{\mu\nu} - \gamma_\mu q_\nu)\,O^{\nu\alpha}(X),
\qquad\qquad\qquad
\overline{E}^{\alpha}_\mu
=
(\not\hskip-0.7mm\!{k}\,g_{\mu\nu} - \gamma_\mu k_\nu)\,O^{\nu\alpha}(X),
\nonumber\\
{W}^{\alpha}_\mu
&=&
O^{\alpha\nu}(X)\,(\not\hskip-0.7mm\!{q}\,g_{\mu\nu} - \gamma_\mu q_\nu),
\qquad\qquad\qquad
{E}^{\alpha}_\mu
=
O^{\alpha\nu}(X)\,(\not\hskip-0.7mm\!{k}\,g_{\mu\nu} - \gamma_\mu k_\nu).
\end{eqnarray}
The resonance propagators ${\cal P}(N^*)$ in
Eqs.(\ref{A1/2})-(\ref{A7/2}) are defined by making use of the
conventional prescription~\cite{IZ} that the  following spectral
decomposition of the $N^*$ field is valid
\begin{eqnarray}
\psi_{N^*}(x)=\int \frac{d^3{\bf p} }{(2\pi)^3\sqrt{2E_p}}
\left[
a_{{\bf p},r} \, u_{N^*}^r(p)e^{-ipx} +
b^{+}_{{\bf p},r} \, v_{N^*}^r(p)e^{+ipx}
\right],
\end{eqnarray}
where $u_{N^*}, v_{N^*}$ are the Rarita-Schwinger spinors.
The finite decay width $\Gamma_{N^*}$ is introduced into
the denominators of the propagators by the substitution
$M_{N^*} \to M_{N^*}  - \frac i2 \Gamma_{N^*}$. Explicitly, we then
have for the $J=\frac{1}{2}$ case(Eq.(31))
\begin{eqnarray}
{\cal P}^{\pm}(N^*)=\frac{\Lambda^{\pm}(p,M_{N^*})}
{p_r^2-M_{N^*}^2 + i\Gamma_{N^*}M_{N^*} },
\end{eqnarray}
The the spin projection operators  $\Lambda^{\pm}({p},M)$
in the above equation  are defined by the bilinear
combinations of the Rarita-Schinger spinors
\begin{eqnarray}
\Lambda^+({p},M)
& = &
\frac12 \sum_r \left(
(1+\frac{p_0}{E_0})u^r({\bf p},E_0)\otimes\bar u^r({\bf p},E_0)
\right. \nonumber \\
& &  \left. \hspace*{10.5mm}
-(1-\frac{p_0}{E_0})v^r({-\bf p},E_0)\otimes\bar v^r({-\bf p},E_0)
\right)\nonumber \\
& = &
 \, \not\hskip-0.7mm\!{p}\,+\,M, \nonumber \\
\Lambda^-({p},M) &=&  \, \not\hskip-0.7mm\!{p}\, - \,M
\label{L1/2}
\end{eqnarray}
where $E_0=\sqrt{{\bf p}^2+M^2}$,
 $u$ and $v$ are the usual Dirac spinors.
 The projection operators $P_{ab}$ with
 $a=\alpha,\alpha\beta,\alpha\beta\gamma$, in
 Eqs.(\ref{A3/2}) - (\ref{A7/2}) take the same form of Eq.(37), except
 that their projection operators have the following higher-rank
 tensor forms
\begin{eqnarray}
\Lambda_{\alpha\beta}({p},M)
& = &
\frac12 \sum_r \left(
(1+\frac{p_0}{E_0}){\cal U}^r_\alpha({\bf p},E_0)
\otimes\bar {\cal U}^r_{\beta}({\bf p},E_0)
\right. \nonumber \\
& &  \left. \hspace*{10.5mm}
-(1-\frac{p_0}{E_0}){\cal V}^r_\alpha({-\bf p},E_0)
\otimes\bar {\cal V}^r_\beta({-\bf p},E_0)
\right);
\label{L3/2}\\
\Lambda_{\alpha\beta,\gamma\delta}({p},M)
& = &
\frac12 \sum_r \left(
(1+\frac{p_0}{E_0}){\cal U}^r_{\alpha\beta}({\bf p},E_0)
\otimes\bar {\cal U}^r_{\gamma\delta}({\bf p},E_0)
\right. \nonumber \\
& & \left. \hspace*{10.5mm} -(1-\frac{p_0}{E_0}){\cal
V}^r_{\alpha\beta}({-\bf p},E_0) \otimes\bar {\cal
V}^r_{\gamma\delta}({-\bf p},E_0) \right),
 \label{L5/2}\\
 \Lambda_{\alpha\beta\gamma,\delta\sigma\xi}({p},M) & = & \frac12 \sum_r
 \left( (1+\frac{p_0}{E_0}){\cal U}^r_{\alpha\beta\gamma}({\bf p},E_0)
 \otimes\bar {\cal U}^r_{\delta\sigma\xi}({\bf p},E_0)
 \right. \nonumber \\
 & & \left. \hspace*{10.5mm} -(1-\frac{p_0}{E_0}){\cal
 V}^r_{\alpha\beta\gamma}({-\bf p},E_0) \otimes\bar {\cal
 V}^r_{\delta\sigma\xi}({-\bf p},E_0) \right), \label{L7/2}
\end{eqnarray}
 where the Rarita-Schwinger spinors are defined by
\begin{eqnarray}
 {\cal U}^r_\alpha(p)
 & = &
 \sum_{\lambda, s} \langle 1\, \lambda \,\frac12\,s|\,\frac32\,
 r \rangle \, \varepsilon^\lambda_\alpha(p) \, u^s(p),\nonumber\\
 {\cal U}^r_{\alpha\beta}(p)
 & = &
 \sum_{\lambda, \lambda' s, t} \langle 1
 \, \lambda \,\frac12 \, s| \, \frac32 \, t \rangle \,
 \langle \frac32 \, t \, 1\, \lambda' | \, \frac52\, r \rangle\,
 \varepsilon^\lambda_\alpha(p) \,
 \varepsilon^{\lambda'}_\beta(p) \,u^s(p),\nonumber\\
 {\cal U}^r_{\alpha\beta\gamma}(p)
 & = &
 \sum_{\lambda, \lambda',\lambda'', s, t,w} \langle 1
 \, \lambda \,\frac12 \, s| \, \frac32 \, t \rangle \,
 \langle \frac32 \, t \, 1\, \lambda' | \, \frac52\, w \rangle\,
 \langle \frac52 \, w \, 1\, \lambda'' | \, \frac72\, r \rangle\,
 \varepsilon^\lambda_\alpha(p) \,
 \varepsilon^{\lambda'}_\beta(p) \,
 \varepsilon^{\lambda''}_\gamma(p) \,
 u^s(p).
 \end{eqnarray}
The spinors $v$ and ${\cal V}$ are related to $u$ and ${\cal U}$ as
$v(p) = i \gamma_2\, u^*(p)$ and
${\cal V}(p)=i\gamma_2\, {\cal U}^*(p)$, respectively.
The polarization four-vector   $\varepsilon^\lambda_\mu$
for a spin-1 particle with spin projection
$\lambda$, four-momentum $p=(E,{\bf p})$ and mass $m$, is
\begin{eqnarray}
\varepsilon^\lambda(p) = \left( \, \frac{{\bm
\epsilon}^\lambda\cdot{\bf p}}{m},\,\, {\bm\epsilon}^\lambda +
\frac{{\bf p}\,({\bm \epsilon}^\lambda\cdot{\bf p})}{m ( E + m)}\,
\right),
\end{eqnarray}
where the three-dimensional polarization vector $\bm \epsilon$ is
defined by
\begin{eqnarray}
{\bm \epsilon}^{\pm 1} = \mp \frac{1}{\sqrt2} (\,1,\,\,\pm
i,\,\,0\,), {\bm \epsilon}^{0} = (\,0,\,\,0,\,\,1\,).
\end{eqnarray}

At first glance it seems that the spin projection
operator Eq.(\ref{L3/2}) for $J=\frac{3}{2}$ $N^*$
is rather different
from the
commonly used on-shell operator
$\overline{\Lambda}_{\alpha\beta}(p,M)$, which
may be obtained~\cite{Pilkuhn} from the Rarita-Schwinger
spinors
\begin{eqnarray}
\overline{\Lambda}_{\alpha\beta}({p},M)
& = &
\sum_{r}{\cal U}^r_\alpha({\bf p})
\otimes\bar {\cal U}^r_{\beta}({\bf p})
\nonumber \\
& = &
-\left[g_{\alpha\beta}-\frac13\gamma_\alpha\gamma_\beta
       -\frac{\gamma_\alpha p_\beta-\gamma_\beta p_\alpha}{3M}
       -\frac{2p_\alpha p_\beta}{3M^2}
\right]( \, \not\hskip-0.7mm\!{p}\,+\,M).
\label{C1}
\end{eqnarray}
We now note that at the on-shell $p^2=M^2$ point, the contribution
from of the  second terms in Eqs.(\ref{L3/2})
vanishes and
${\Lambda}_{\alpha\beta}(p,M)=\overline{\Lambda}_{\alpha\beta}(p,M)$.

There are some arbitrariness in defining the propagators of higher
spin particles. In this work we use the prescription based on
Eqs.(\ref{A1/2})- (\ref{A5/2}) and Eqs.(\ref{L1/2}) - (\ref{L5/2})
because (i) it  coincides at mass shell with the well known spin
$\frac32$ propagator Eq.(\ref{C1}) and (ii) it gives correct
off-shell behavior for $\frac12$ baryons. We remark
 that the form Eq.(\ref{C1}) for
$J=\frac{3}{2}$ particles
 has some unphysical features for the
off-shell $p^2 \neq M^2$ cases,
which can not be fixed for both the $s$ and $u$ channel $N^*$
excitations by making the simple substitution[39] $M \rightarrow
\sqrt{p^2}$.

The effect of the finite resonance decay width is quite different
for $s$ and $u$ channels, because of the evident relation $|u|+M^2_{N^*} \gg
|s-M^2_{N^*}|$. Therefore, for simplicity, for the $u$ channels
we use a constant value $\Gamma_{N^*} = \Gamma^0_{N^*}$.
 For the $s$ channels the energy-dependent widths are
calculated according to Ref.\cite{MS92}
\begin{eqnarray}
\Gamma(W) = \sum_{j}\Gamma_{j}\frac{\rho_{j}(W)}{\rho_j(M_{N^*})}.
\label{B1}
\end{eqnarray}
where $\Gamma_{j}$ is the partial width for the resonance decay
into the $j$th channel, evaluated at $W \equiv\sqrt{s}=M_{N^*}$.
The form of the "space-phase" factor $\rho_j(W)$ depends on the
decay channel, the relative momentum $q_j$ of the outgoing particles,
 and their relative orbital momentum $l_j$. It provides  proper analytic
threshold behaviour $\rho_{j}(W)\sim q_j^{2l_j+1}$
as  $q_j\to 0$ and becomes constant at high energy~\cite{MAGT84}.

Our next task is to fix the coupling constants in
the  effective Lagrangians Eqs.(\ref{R1/2}),(\ref{R3/2}),
(\ref{R5/2}), and (\ref{R7/2}).
They can be calculated by using the empirical helicity amplitudes
listed by Particle Data Group~\cite{PDG98}.
For the spin $J=\frac{1}{2}$ resonances we find that
\begin{eqnarray}
eg_{a}
 =
\pm \frac{C}{2\sqrt{2}k^*}\,A^{\frac12}_{a},
 \label{eg1/2}
\end{eqnarray}
with
 \begin{eqnarray}
 k^* = \frac{M_{N^*}^2 - M_N^2}{2M_{N^*}},\qquad C=\sqrt{8M_NM_{N^*}k^*}
 \label{k}
 \end{eqnarray}
and $a=p,n,s$.

 For spin $J= \frac32,\,\frac52,\,\frac72$ particles,
 the  relations between the helicity amplitudes and
 coupling constants are given by
\begin{eqnarray}
  J^P={\frac{3}{2}}^{\pm}~~~&&\nonumber\\
  eg^{}_{a}& =& \pm \frac{\sqrt{3}C}{2M_N}\frac{M_{N^*}}{k^*}\,A^{\frac12}_{a},
  \qquad eg^{}_{a} = \pm \frac{C}{2M_{N^*}}\frac{M_{N^*}}{k^*}
  \,A^{\frac32}_{a},
  \label{eg3/5II}\\
 J^P={\frac{5}{2}}^{\pm}~~~&&\nonumber\\
  eg^{}_{a}& =& \mp \frac{\sqrt{5}C}{2M_N}
  \left(\frac{M_{N^*}}{k^*}\right)^2\,A^{\frac12}_{a},
  \qquad
  eg^{}_{a}= \mp \frac{\sqrt{5}C}{2\sqrt{2}M_{N^*}}
  \left(\frac{M_{N^*}}{k^*}\right)^2
  \,A^{\frac32}_{a},
 \label{eg5/2II}\\
  J^P={\frac{7}{2}}^{\pm}~~~&&\nonumber\\
  eg^{}_{a} &=& \pm \frac{\sqrt{35}C}{4M_N}
  \left(\frac{M_{N^*}}{k^*}\right)^3\,A^{\frac12}_{a},
  \qquad
  eg^{}_{a}= \pm \frac{\sqrt{21}C}{4M_{N^*}}
  \left(\frac{M_{N^*}}{k^*}\right)^3
  \,A^{\frac32}_{a},
 \label{eg7/2II}
 \end{eqnarray}
  The above equations lead to the relations:
  $A^{3/2}/A^{1/2}=\sqrt{3}M_{N^*}/M_N$,
  $(\sqrt{2}M_{N^*}/M_N)$ and $(\sqrt{5}M_{N^*}/\sqrt{3}M_N)$,
  for the resonances
  with $J^P=\frac32^\pm\,\,(\frac52^\pm)$ and $(\frac72^\pm)$,
  respectively.  Since this strong
  correlation is not observed for all resonances, we evaluate
  $g^{}$ by normalizing it to
  $(A^{1/2}_a)^2 + (A^{3/2}_a)^2$ and
  taking it's sign to be the  sign
  of the dominant component. Instead of using the above two
  equations, we therefore  calculate the coupling constants for
  $J=\frac{3}{2}, \frac{5}{2},$ and $\frac{7}{2}$ particles by using the following
  relations:
 \begin{eqnarray}
  J^P={\frac{3}{2}}^{\pm}~~~&&\nonumber\\
  eg^{}_{a}& =&
  \pm S^{\frac32}_a
  \frac{\sqrt{3}\,C}{2\sqrt{3M_{N^*}^2+M_N^2}}\frac{M_{N^*}}{k^*}
  \left((A^{\frac12}_a)^2 + (A^{\frac32}_a)^2
  \right)^\frac{1}{2},
  \nonumber\\
  S^{\frac32}_a&=&{\rm sign}(A^{\frac12})\theta(h_a) +
  {\rm sign}(A^{\frac32})\theta(-h_a),
  \qquad h_a=|A^{\frac12}_{a}| -
  \frac{1}{\sqrt{3}}\frac{M_N}{M_{N^*}}|A^{\frac32}_{a}|;
  \label{eg3/2II-}\\
  J^P={\frac{5}{2}}^{\pm}~~~&&\nonumber\\
  eg^{}_{a}& =&
  \mp S^{\frac52}_a
  \frac{\sqrt{5}\,C}{2\sqrt{2M_{N^*}^2+M_N^2}}
  \left(\frac{M_{N^*}}{k^*}\right)^2
  \left((A^{\frac12}_a)^2 + (A^{\frac32}_a)^2
  \right)^\frac{1}{2},\nonumber\\
  S^{\frac52}_a&=&{\rm sign}(A^{\frac12})\theta(h_a) +
  {\rm sign}(A^{\frac32})\theta(-h_a),
  \qquad h_a=|A^{\frac12}_{a}| -
  \frac{1}{\sqrt{2}}\frac{M_N}{M_{N^*}}|A^{\frac32}_{a}|.
  \label{eg5/2II-}\\
 J^P={\frac{7}{2}}^{\pm}~~~&&\nonumber\\
   eg^{}_{a}& =&
  \pm S^{\frac72}_a
  \frac{\sqrt{105}\,C}{4\sqrt{5M_{N^*}^2+3M_N^2}}
  \left(\frac{M_{N^*}}{k^*}\right)^2
  \left((A^{\frac12}_a)^2 + (A^{\frac32}_a)^2
  \right)^\frac{1}{2},\nonumber\\
  S^{\frac52}_a&=&{\rm sign}(A^{\frac12})\theta(h_a) +
  {\rm sign}(A^{\frac32})\theta(-h_a),
  \qquad h_a=|A^{\frac12}_{a}| -
  \sqrt{\frac{3}{5}}\frac{M_N}{M_{N^*}}|A^{\frac32}_{a}|.
  \label{eg7/2II-}
  \end{eqnarray}
In above equations, we have defined $\theta(a)=1, (0)$ for $a >(<)
0$ and ${\rm sign}(A)= A/\mid A \mid$ denoting the sign of $A$.
Obviously, our procedure can be used to investigate the $N^*$
excitation in $\gamma N \rightarrow \omega N$ reaction only for
the resonances with $\gamma N \rightarrow N^*$ helicity amplitudes
given by PDG.

\section{Results and Discussion}

To perform calculations, the parameters of the
$\gamma N \rightarrow \omega N$ amplitudes
 defined in Sections II and III must be specified.
As discussed  in
section II, the parameters of the
non-resonant amplitudes are  taken from our
previous study of $\gamma p \rightarrow \omega p$ reaction at higher
 energies~\cite{OTL01}. We therefore only need to
choose the parameters of the resonant amplitudes developed in
Sec.~III. These parameters are: the coupling constants $g_{\gamma
NN^*}$ and $f_{\omega NN^*}$ in ~Eqs.~(\ref{A1/2}) - (\ref{A7/2}),
the cut-off parameters $\Lambda_{N^*}$ for the $\gamma N N^*$ and
$\omega N N^*$  form factors of the form of Eq.~(\ref{cutN}), and
the off-shell parameters $Z$ of Eq.(23) for the higher spin
resonances.

The coupling constants, $g_{\gamma NN^*}$,  are obtained
by using the empirical helicity amplitudes listed by
the Particle Data Group(PDG) \cite{PDG98} to evaluate
Eqs.~(\ref{vdm2}), (\ref{vdm3}), (\ref{eg1/2}), (\ref{eg3/2II-}) -
(\ref{eg7/2II-}). The corresponding $f_{\omega N N^*}$
constants are then obtained by using the relations (27)-(29)
of the vector meson dominance model.
 Because of the uncertainties in the
empirical helicity amplitudes, we consider two cases(models). In
the model I we use the central values of PDG's  helicity
amplitudes $A_{p(n)}^{\frac12}$, $A_{p(n)}^{\frac32}$. These
empirical helicity amplitudes for the considered $N^*$'s  along
with the calculated $\gamma N N^*$ and $\omega N N^*$ coupling
constants are listed in Table I. Note that  some high mass
resonances listed by PDG are not included in Table I. These
resonances can not be included in our investigation because of
their $\gamma N \rightarrow N^*$ helicity amplitudes are not
available. These resonances are expected to have negligible
effects in the near threshold region.

Concerning the off-shell parameter $Z$, we take the simplest
approach by setting
$Z=-1/2$ such that the second term of Eq.(\ref{O_munu})
 does not contribute.
This is a rather arbitrary choice, but is supported by our finding
that the calculated amplitude is rather insensitive to $Z$ in a
very wide range of $-5 \le Z \le +5$. This is due to the fact that
the dominant $N^*$ contribution comes from the $s$-channel
diagram(Fig.1d) for which
$\gamma^\alpha\Lambda_{\alpha\cdot\cdot}(p_s,M_{N^*})\simeq 0$ for
all projection operators $\Lambda$ defined in Eqs.(35)-(41). As a
result, the contributions from the $Z$-dependent terms in
Eq.(\ref{O_munu}), which are  proportional to
$\gamma_\mu\gamma_\nu$, are very small.

With the above specifications, the only free parameters of model I
are the cut-off $\Lambda_{N^*_i}$ for each of resonances. For
simplicity, we assume the same cutoff for all $N^*$'s and set
$\Lambda_{N^*_i}= \Lambda_{N^*}$. We then find that the
unpolarized differential cross sections at low energies can be
best described by setting $\Lambda_{N^*}$=0.7 GeV/c. The result is
the solid curve shown in the left-side of Fig.2. The predicted
beam asymmetry is the dashed curve in Fig.3. Obviously, the model
I does not give a good account of the data displayed in Figs. 2
and 3.

The use of the central values of the empirical helicity amplitudes
as input to our calculations is perhaps too restrictive.
We therefore consider model II by allowing the values of
$\gamma N N^*$ and $\omega N N^*$ coupling constants to vary
 within the
ranges allowed by the uncertainties of the empirical helicity
amplitudes. We then find that it is possible to get a good
description of the existing  data of both the unpolarized
differential cross section at $E_\gamma=1.23$
GeV~\cite{Klein96-98} and the beam asymmetry at $E_\gamma=1.175$
GeV~\cite{GRAAL}. The results are the solid curve in the
right-hand-side of Fig.2 and the dot-dashed curve in Fig.3. The
resulting parameters which are different from those of model I are
listed in Table II. The cutoff used in this fit is
$\Lambda_{N^*}=0.75$ GeV/c.

 In Fig.~\ref{fig:2} we also show the contributions from the
non-resonant amplitude (dot-dashed lines) and from the resonance
excitation (dashed line). We see that the resonant contributions
in two models have rather different $t$-dependent. The much larger
resonant contribution in model II(right-hand-side of
Fig.\ref{fig:2} clearly is essential in obtaining the agreement
with the data. The parameters of model II, listed in Table II, are
very suggestive in future determination of the $N^*$ parameters.
We have also found that the dominant contribution to the
 non-resonant amplitude comes from the pseudoscalar exchange.
This is consistent with our finding\cite{OTL01} in the investigation
at higher energies.

 More sizeable  differences between the constructed two models
can be seen in Fig.~\ref{fig:3} for the beam asymmetry
at $E_\gamma=1.175$ GeV.  The
 non-resonant amplitude alone yields  zero asymmetry. The
 negative asymmetry is due to the interference between the
non-resonant and resonant amplitudes. We find that the dominant
resonant contribution in
 both models comes from the excitation of $F_{15}(1680)$  state.
 To illustrate
 this, we show in Fig.~\ref{fig:4} the beam asymmetry calculated
 from keeping only the $F_{15}(1680)$ term in the resonant
 amplitude. We see that the main feature of the beam asymmetry can
 be obtained( dashed and dot-dashed curves) from the interference
 between this resonance state and the non-resonant amplitude.  The
 next large contribution comes from the excitation of
 $D_{13}(1520)$, $S_{11}(1650)$ and $P_{13}(1720)$
 states. Their role in determining the beam asymmetry is rather
 different.
 The $D_{13}(1520)$  resonance gives constructive
 interference with
 $F_{15}(1680)$, while the $S_{11}(1650),\,P_{13}(1720)$ excitations
 interfere destructively  and reduce the absolute value of
 asymmetry.

With the model II constructed above,
we perform calculations
 at $E_\gamma =$ $1.125$,  $1.175$, and $1.23$ GeV
for future experimental tests.
The predicted differential cross sections
are the thick solid curves
shown in Fig.~\ref{fig:5}. Here, we also show
 the contributions from
various mechanisms illustrated in Fig.1.
 One can see that the contributions from the direct and crossed
nucleon terms(short dashed curves) and $N^*$ excitations(dashed
curves) are instrumental in getting differential cross
sections which are much flatter than those calculated
 from keeping only the
pseudo scalar meson exchange (thin solid curves).

 The predicted
 beam asymmetries, $\Sigma_x$, are
 shown in Fig.~\ref{fig:6}.  The calculations keeping only the
 non-resonant amplitude
 yield almost zero asymmetries(dot-dashed curves).
 Inclusion of the resonance
 excitation results in negative asymmetry, in agreement with
 the data at $E_\gamma =1.125$ and $1.175$ GeV.

 In Figs.~\ref{fig:7},\ref{fig:8} we depict the predicted target
 asymmetry($T_y$) and recoil asymmetry($P_y$).
 Again, one can see that
 resonance excitation results in large deviations
 from the pure non-resonant limit(dot-dashed curves).

In Fig.~\ref{fig:9} we show our predictions for the tensor $V_{z'z'}$
 asymmetry calculated in the Gottfried-Jackson system.
 The dot-dashed curves are calculated from keeping only
 the non-resonant amplitudes. Clearly, these are
  close to the value $1/\sqrt{2}$ of Eq.(\ref{VGJ}) due to only the
 pseudoscalar meson exchange.  The resonance excitations
 enhance greatly
the  contribution from the  longitudinally polarized outgoing
 $\omega$ mesons and hence bring  $V_{z'z'}$ to negative
 values.

 The predicted
double beam-target asymmetry is shown in Fig.~\ref{fig:10}. Here
we see even more dramatic effects due to $N^*$ excitations. The
non-resonant amplitude yields positive asymmetry. Adding the
resonant contributions, the asymmetries at low $E_\gamma$ become
negative and have very  different dependence on scattering angles.

 The results presented above obviously reflect
 the consequences of the not-well-determined
 $\omega N N^*$ coupling constants. It is therefore interesting to
 compare our values with those determined in the recent works by
 Post and Mosel~\cite{PM00}, Lutz, Wolf and Friman~\cite{LWF01},
 and Riska and Brown~\cite{RB00}. The form of the effective Lagrangians
given in Ref.\cite{PM00} can be obtained from our expressions by
making the non-relativistic reductions. In this case the
comparison can be made easily. For other two cases their
 relations with ours are not obvious because of
the use of  different couplings schemes. However, by  making the
non-relativistic reductions and keeping the leading terms which
would be dominant near the $\omega$ production threshold,  we can
cast our Lagrangians into their forms and the comparisons of the
coupling constants can then be made. The resulting  relations are

\begin{eqnarray}
  f_{\omega NN^*}^{J^P=\frac12^+}
  &\Rightarrow&
  \frac{M_{N^*}}{m_\omega}f_{\omega NN^*}^{J^P=\frac12^+}({\rm
Ref.}[20]),\qquad
  \frac{M_{N^*}}{M_N^*+M}g_{\omega NN^*}^{J^P=\frac12^+}({\rm
Ref.}[19]),\nonumber\\
  f_{\omega NN^*}^{J^P=\frac12^-}
  &\Rightarrow&
  \frac{M_{N^*}}{m_\omega}f_{\omega NN^*}^{J^P=\frac12^-}({\rm
Ref.}[20]),\qquad
  \frac{M_{N^*}}{M_N^*-M}g_{\omega NN^*}^{J^P=\frac12^-}({\rm
  Ref.}[19]),\nonumber\\
 &\Rightarrow&
  \frac{M_{N^*}}{\sqrt{3}(M_N^*-M)}f_{\omega NN^*}^{J^P=\frac12^-}({\rm
Ref.}[36]),\nonumber\\
  f_{\omega NN^*}^{J^P=\frac32^+}
  &\Rightarrow&
  \frac{M_{N^*}}{2m_\omega}f_{\omega NN^*}^{J^P=\frac32^+}({\rm
Ref.}[20]),\qquad
  \frac{M_{N^*}}{M_N^*+M}g_{\omega NN^*}^{J^P=\frac32^+}({\rm
Ref.}[19]),\qquad\nonumber\\
  f_{\omega NN^*}^{J^P=\frac32^-}
  &\Rightarrow&
  \frac{M_{N^*}}{m_\omega}f_{\omega NN^*}^{J^P=\frac32^-}({\rm
Ref.}[20]),\qquad
  \frac{M_{N^*}+M_N}{4M}g_{\omega NN^*}^{J^P=\frac32^-}({\rm
  Ref.}[19]),\nonumber\\
  &\Rightarrow&
  \frac{M_{N^*}}{(M_N^*-M)}f_{\omega NN^*}^{J^P=\frac32^-}({\rm
Ref.}[36]),\nonumber\\
  f_{\omega NN^*}^{J^P=\frac52^+}
  &\Rightarrow&
  \frac{M_{N^*}^2}{m_\omega^2}f_{\omega NN^*}^{J^P=\frac32^-}({\rm
Ref.}[20]),\qquad
  \frac{M_{N^*}^2}{am_\omega^2} g_{\omega NN^*}^{J^P=\frac32^-}({\rm
Ref.}[19]),\nonumber\\
  {\rm with}\,\,a^2&\simeq& 1+ \frac{4M^2(M_{N^*}-M)^2}{m_\omega^4} +
        \frac{2M(M_{N^*}-M)^2}{m_\omega^2} + \ldots\,\,\, ,\nonumber\\
 f_{\omega NN^*}^{J^P=\frac52^-}
  &\Rightarrow&
  \frac{M_{N^*}^2}{2m_\omega^2} g_{\omega NN^*}^{J^P=\frac32^-}({\rm
  Ref.}[19]),
 \end{eqnarray}
 where the factor $a$ accounts the enhancement of transversely
 polarized $\omega$ in our model.

The comparison is given in Table III. We see that there are some
reasonable agreements between the considered four approaches. On
the other hand, very  large differences exist in several cases.
This is not surprising at this stage of development. Only the use
of more experimental data, such as the spin observables discussed
in this work, can improve the situation.

\section{Summary}

In summary, we have investigated the role of
nucleon resonances in $\omega$ photoproduction
 in the near threshold energy region by using
the effective Lagrangian approach and the vector
dominance model. By using the empirical helicity amplitudes
of $\gamma N \rightarrow N^*$ transitions, a
phenomenological model( model II) has been obtained to give a
good description of
the existing data of both the differential
cross sections at $E_\gamma=1.23$ GeV and
the beam asymmetry at $E_\gamma =1.125$ and $1.175$ GeV.
We have found that in the near threshold energy region $E\gamma \leq
1.25$ GeV, the dominant resonant
contribution comes from $F_{15}(1680)$ and $D_{13}(1520)$
$N^*$ states. The single and double polarization observables
are strongly modified by the $N^*$ excitations.
It will be interesting to obtain data for testing our
predictions given in Figs.(6)-(10).

To end, we emphasize that the tree-diagram calculations
based on effective Lagrangian approach
is known to be valid only in the energy region close to
threshold. For investigating $N^*$ effects at higher energies,
it is necessary to include the  initial and final
state interactions. A dynamical approach, similar to what has been
developed\cite{SL96} for pion photoproduction, may have to be
developed.

\acknowledgments

This work was supported in part by U.S. Department of Energy,
Nuclear Physics Division, Contract No. W-31-109-ENG-38. A.I.T.
thanks the warm hospitality of the nuclear theory group at Argonne
National Laboratory.

\begin{table}
  \caption{ The helicity amplitudes
 $A^\lambda_{p(n)}$ and coupling constants for effective
 Lagrangians of Eqs.~(\ref{R1/2}) - (\ref{R7/2}) for the model~I.
 The helicity amplitude $A^\lambda_{p(n)}$ is given in unit of
 $10^{-3}$ GeV$^{-1/2}$.
 The resonance mass $M_{N^*}$ is in unit of
 MeV.
 For the heavy resonances $F_{17}(1990)$ and  $D_{13}(2080)$
 we use the data of [Awajii 81]  and for
 $G_{17}(2190)$ - the data of [Crawfold 80],
 listed in \protect\cite{PDG98}.
 The resonance mass $M_{N^*}$ is in unit of
 MeV.
 The helicity amplitude $A^\lambda_{p(n)}$ is given in unit of
 $10^{-3}$ GeV$^{-1/2}$.}
\label{tab:I}
\begin{ruledtabular}
\begin{tabular}{rrrrrrrrrrr}
$N^*$& $M_{N^*}$ & $A^{1/2}_p$ & $A^{1/2}_n$ & $eg_p$  &$f_\omega$
&  \\ \hline
$P_{11}$ & $1440$ & $-65\pm4$   & $40\pm10$  & $-0.117$ & $-1.271$ &\\
$S_{11}$ & $1535$ & $90\pm30$   & $-46\pm27$  & $-0.156$ & $-2.144$ &\\
$S_{11}$ & $1650$ & $53\pm16$   & $-15\pm21$  & $-0.088$ & $-1.782$ &\\
$P_{11}$ & $1710$ & $22\pm16$   & $-2\pm14$  & $0.015$ & $0.323$ &\\
\hline $N^*$& $M_{N^*}$ & $A^{1/2}_p$ & $A^{1/2}_n$ & $A^{3/2}_p$
&$A^{3/2}_n$ & ${eg}_p$ & $f_\omega$ & \\ \hline $D_{13}$ & $1520$
& $-24\pm9$ & $-59\pm9$  & $166\pm5$ & $-139\pm11$ &
$-0.389$ & $5.70$ & \\
$D_{13}$ & $1700$ & $-18\pm13$ & $0\pm50$  & $-2\pm20$ & $-3\pm40$
&
$0.04$  & $1.164$ & \\
$P_{13}$ & $1720$ & $18\pm30$ & $1\pm15$  & $-19\pm20$ &
$-29\pm20$ &
$0.058$ & $3.20$ & \\
 $D_{13}$ & $2080$ & $-20\pm8$ & $7\pm13$  & $17\pm11$ & $-53\pm34$
 & $0.055$ & $2.264$ & \\
$D_{15}$ & $1675$ & $19\pm8$ & $-43\pm12$  & $15\pm9$ & $-58\pm13$
&
$0.242$ & $-13.89$ & \\
$F_{15}$ & $1680$ & $-15\pm6$ & $29\pm9$  & $133\pm12$ & $-33\pm9$
&
$-1.33$ & $-28.38$ & \\
 $F_{17}$ & $1990$ & $30\pm29$ & $-1 $  & $86\pm60$ &
 $-178$ &
 $2.85$ & $-84.92$ & \\
 $G_{17}$ & $2190$ & $-55 $ & $-42 $  & $81\pm12$ & $-126\pm9$
 & $2.74$ & $84.23$ & \\
\end{tabular}
\end{ruledtabular}
\end{table}


 \begin{table}
 \caption{ The helicity amplitudes $A^\lambda$ and
 coupling constants for the model~II which values are different
 from the corresponding values listed in the Table~I.
 Notation is the same as in Table~\ref{tab:I}.}
 \label{tab:II}
 \begin{ruledtabular}
 \begin{tabular}{rrrrrrrrrrr}
 $N^*$& $M_{N^*}$ & $A^{1/2}_p$ & $A^{1/2}_n$ & $eg_p$  &$f_\omega$
 &  \\ \hline
 $S_{11}$ & $1650$ & $37 $   & $-36 $  & $-0.062$ & $-0.047$ &\\
 \hline $N^*$& $M_{N^*}$ & $A^{1/2}_p$ & $A^{1/2}_n$ & $A^{3/2}_p$
 &$A^{3/2}_n$ & ${eg}_p$ & $f_\omega$ & \\ \hline
 $P_{13}$ & $1720$ & $48$ & $-14$  & $-39$ & $-49$ &
 $0.117$ & $-5.63$ & \\
 $F_{15}$ & $1680$ & $-9$ & $38$  & $145$ & $-24$ &
 $-1.45$ & $-34.97$ & \\
\end{tabular}
\end{ruledtabular}
 \vspace*{0.5cm}
\end{table}


 \begin{table}
  \caption{ Comparison of the absolute values of $f_{\omega NN^*}$ coupling
  constants in this model
  and the previous analysis by Riska and Brown~[19],
  Post and Mosel~[20], Luts, Wolf and Friman~[36].}
  \label{tab:III}
 \begin{ruledtabular}
 \begin{tabular}{ccccccccccc}
 $$&        $N^*(1440)$&$N^*(1520)$& $N^*(1535)$&$N^*(1650)$&$N^*(1680)$&$N^*(1675)$&$N^*(1700)$&$N^*(1710)$&$N^*(1720)$
 &  \\ \hline
 ${\rm this}\atop{\rm work} $    & $1.27$&       $5.70$    & $2.14$  &$0.047$ &$34.97$    &$13.89$  &$1.16$ &$0.323$ &$5.63$
 &  \\ \hline
  [19]
               &$3.33$         &$5.04$&   $11.57$   &$5.1$          &$29.5$       &$21.56$     &$1.69$   &$-$&$1.74$
 &  \\ \hline
 [20]
          &$1.6\pm0.9$  &$6.6\pm1.7$  & $1.5\pm2.4$&$2.4\pm 2.3$&$30.1\pm 6.9$ &$-$   &$0.4\pm2.2$&$0.44\pm2.23$&$1.97\pm3.5$
 &  \\ \hline
 [36]&$-$&$11.4$&$2.8$ &$2.9$    &$-$&$-$&$-$&$-$&$-$
 &  \\ \hline
\end{tabular}
\end{ruledtabular}
 \vspace*{0.5cm}
\end{table}

\begin{figure}
\centering \epsfig{file=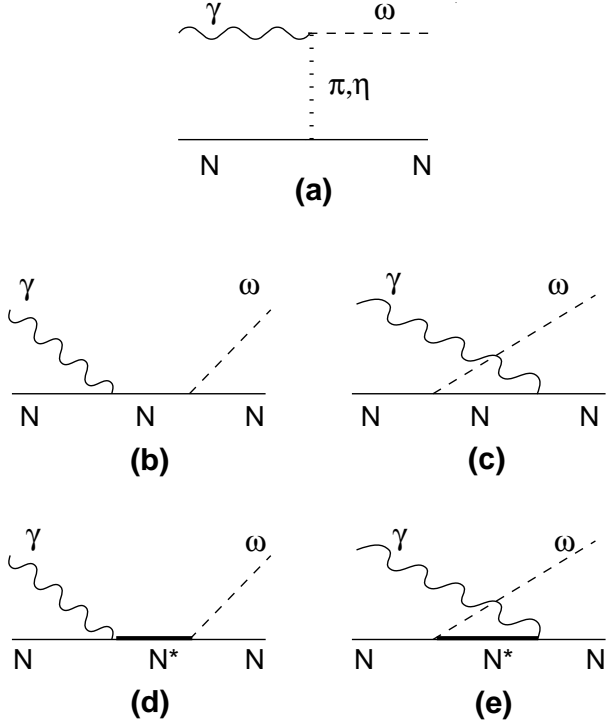, width=8cm} \vspace*{0.5cm}
\caption{ Diagrammatic representation of $\omega$ photoproduction
mechanisms: (a) ($\pi,\eta$) exchange, (b,c) direct and crossed
nucleon terms, (d,e) direct and crossed resonant  terms.}
\label{fig:1}
\end{figure}

\begin{figure}
\centering \epsfig{file=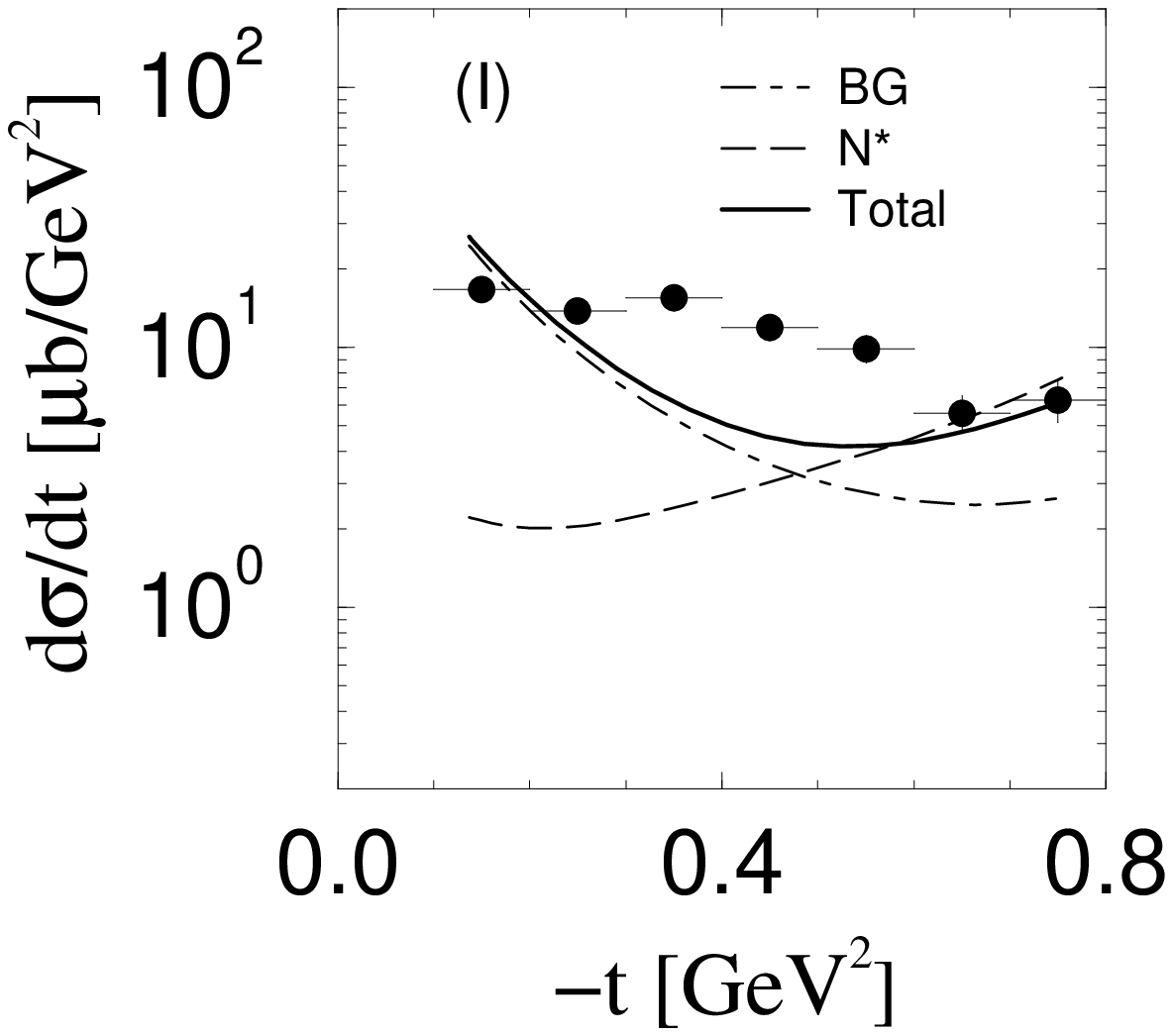, width=4.8cm}\qquad
\epsfig{file=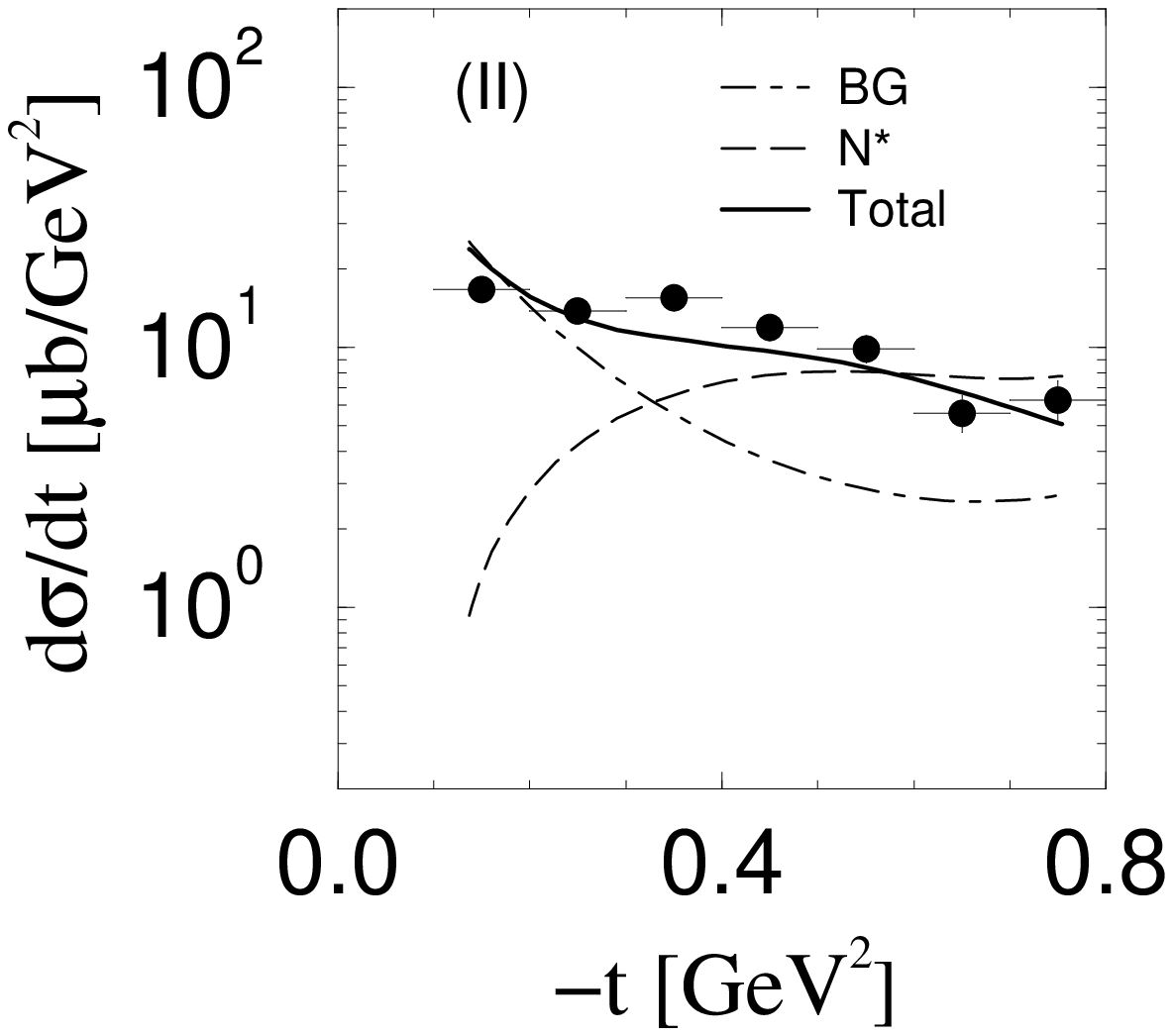,width=4.8cm} \vspace*{0.5cm}
\caption{
Differential cross sections
for $\gamma p\to p\omega$ reaction as a function of $t$ at
$E_\gamma =1.23 $ GeV for the models~I (left panel),
and II (right panel). The results are non-resonant channel
(dot-dashed), resonance  excitation (dashed), and the full
amplitude (solid). Data are taken from Ref.
\protect\cite{Klein96-98}.} \label{fig:2}
\end{figure}

\begin{figure}
\centering
\epsfig{file=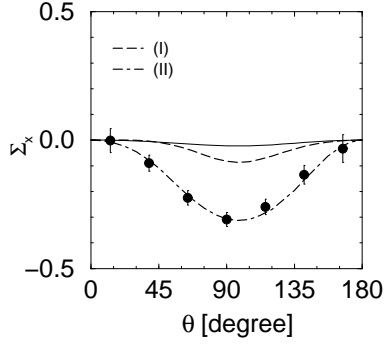, width=5.0cm}
\vspace*{0.5cm}
\caption{
Beam asymmetry as a function of $\omega-$meson production angle
at  $E_\gamma=1.175$ GeV for the two models. The thin solid line
is the beam asymmetry for the non-resonant background, taken separately.
Data are taken from Ref.~\protect\cite{GRAAL}}
\label{fig:3}
\end{figure}

\begin{figure}
\centering
\epsfig{file=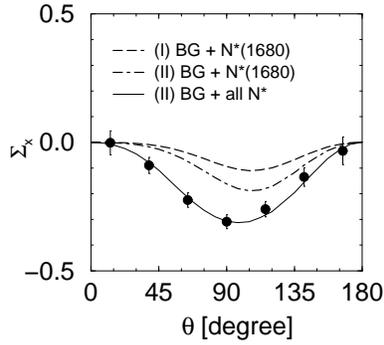, width=5cm}
\vspace*{0.5cm}
\caption{
Beam asymmetry as a function of $\omega$ production angle
at  $E_\gamma=1.175$ GeV for the three models for the coherent sum
of non-resonant background and only $F_{\frac52^+}(1680)$ resonance.
Notation is the same as in Fig.~{\ref{fig:3}} }
\label{fig:4}
\end{figure}

\begin{figure}
\centering \epsfig{file=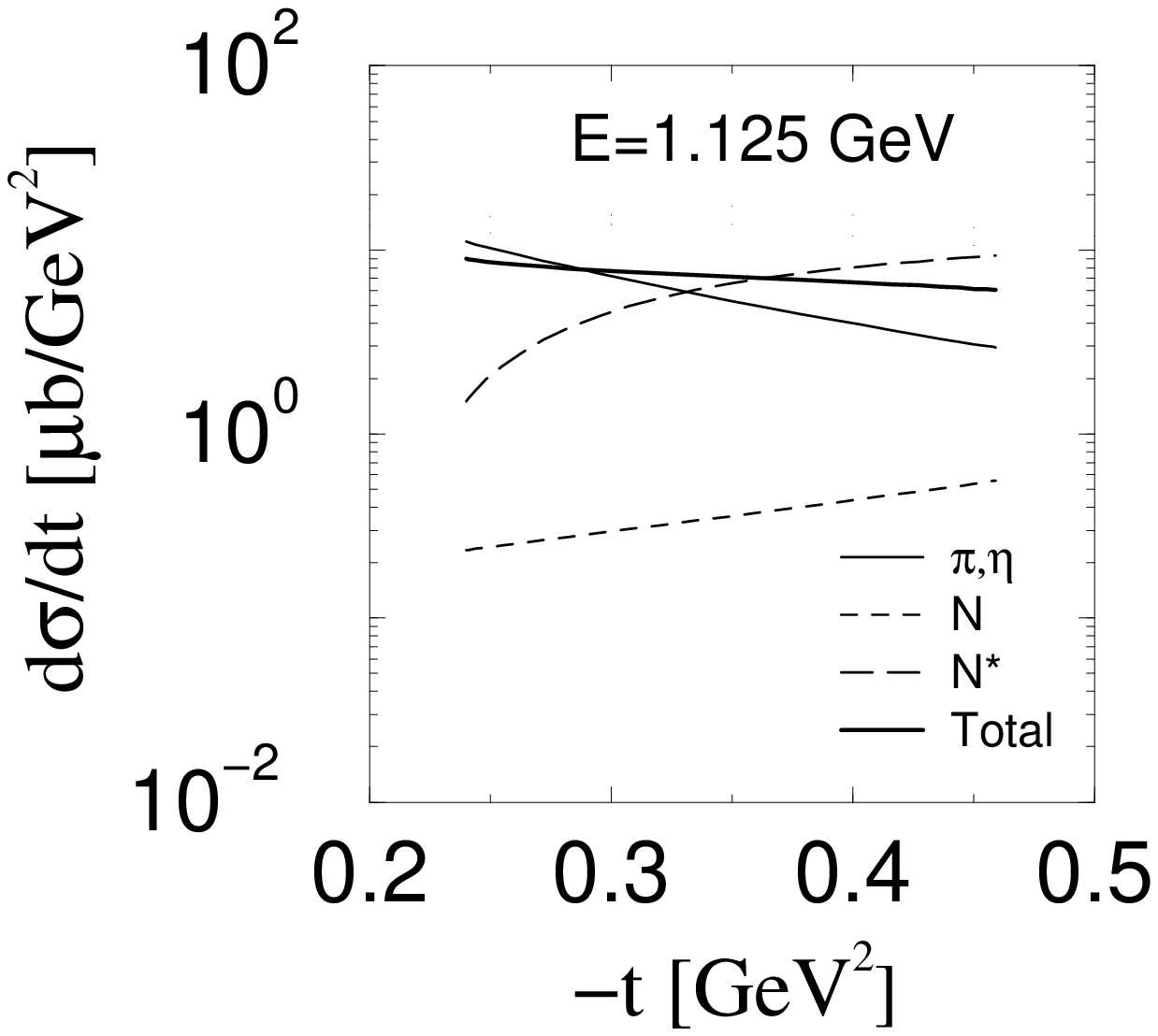, width=4.8cm}\qquad
\epsfig{file=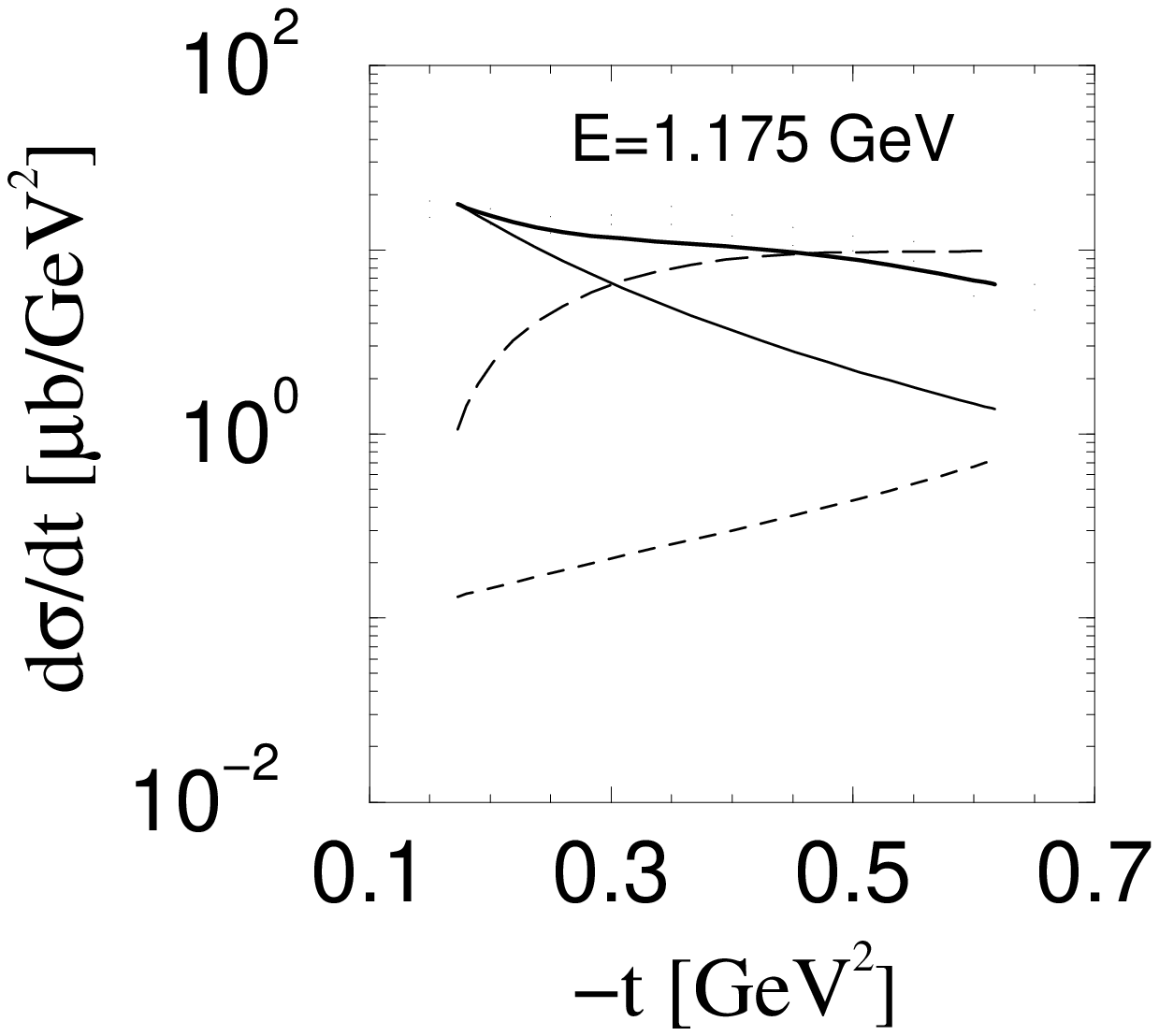, width=4.8cm}\qquad
\epsfig{file=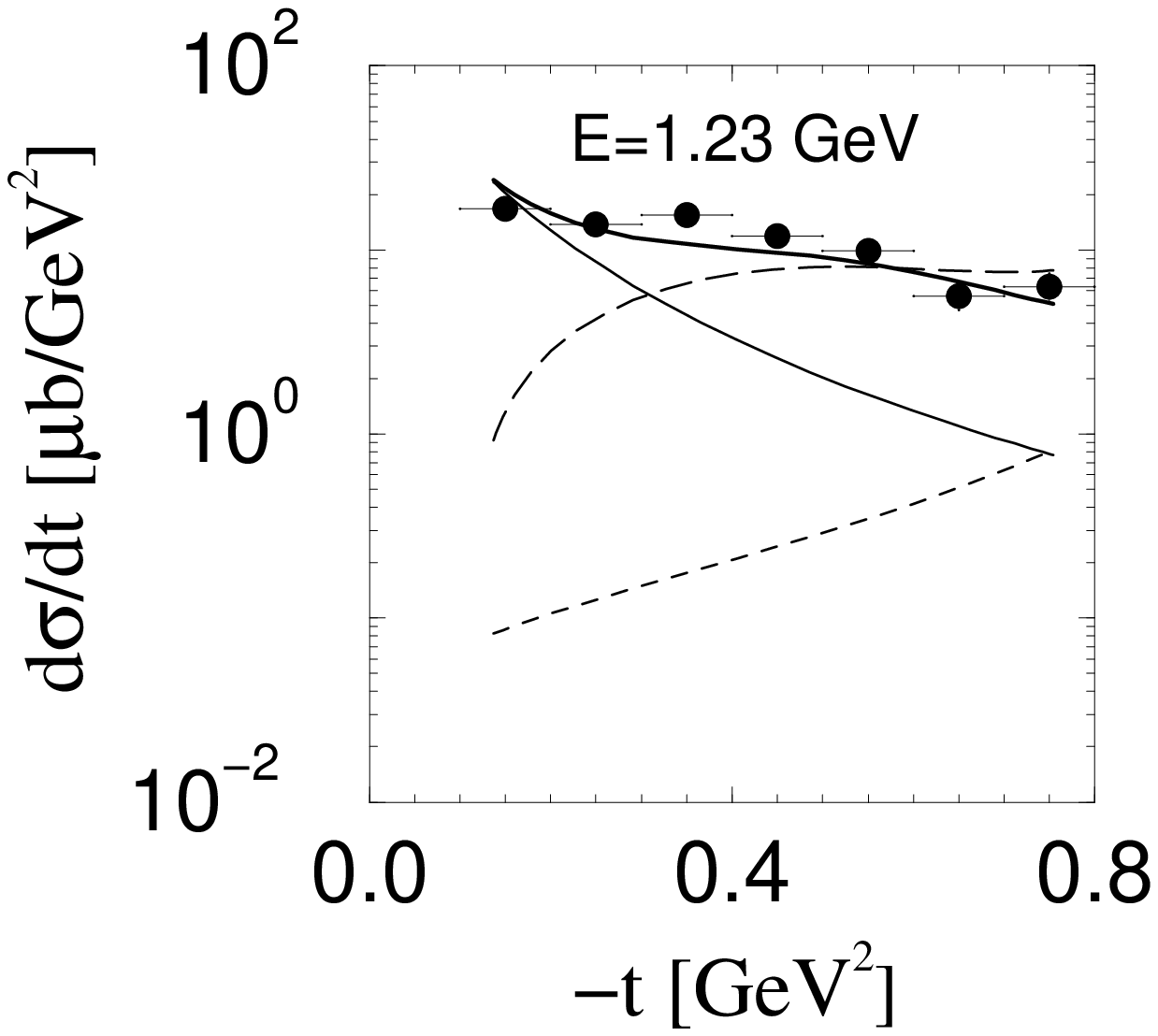, width=48mm} \vspace*{0.5cm} \caption{
Differential cross sections for $\gamma p\to p\omega$ reaction as
a function of $t$ at $E_\gamma =$ $1.125$,  $1.175$, and $1.23$
GeV. The results are from pseudoscalar-meson exchange (thin
solid), direct and crossed nucleon terms (short dashed), $N^*$
excitation (dashed), and the full amplitude (thick solid). Data
are taken from Ref. \protect\cite{Klein96-98}.} \label{fig:5}
\end{figure}

\begin{figure}
 \centering \epsfig{file=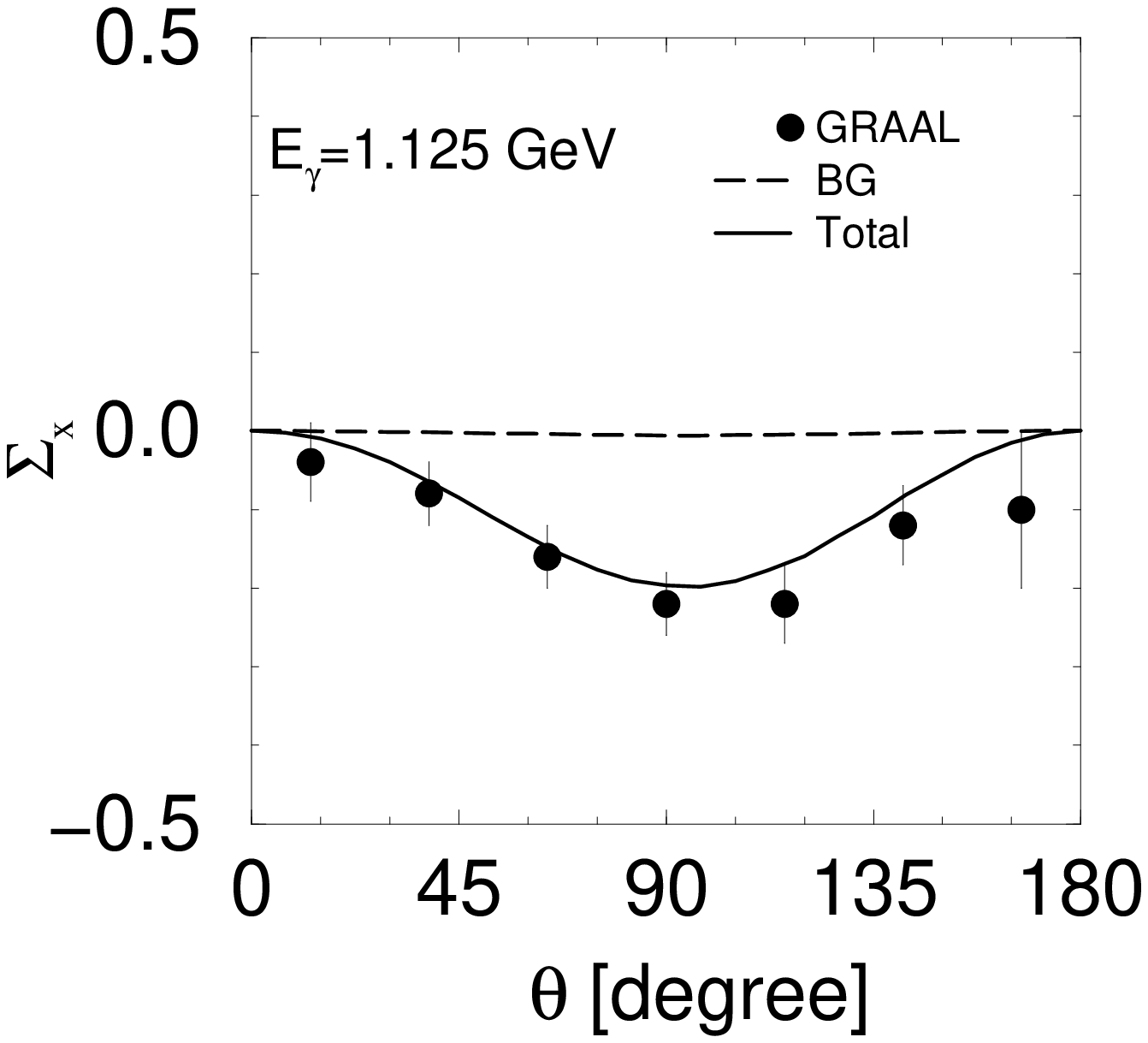, width=46mm}\qquad
 \epsfig{file=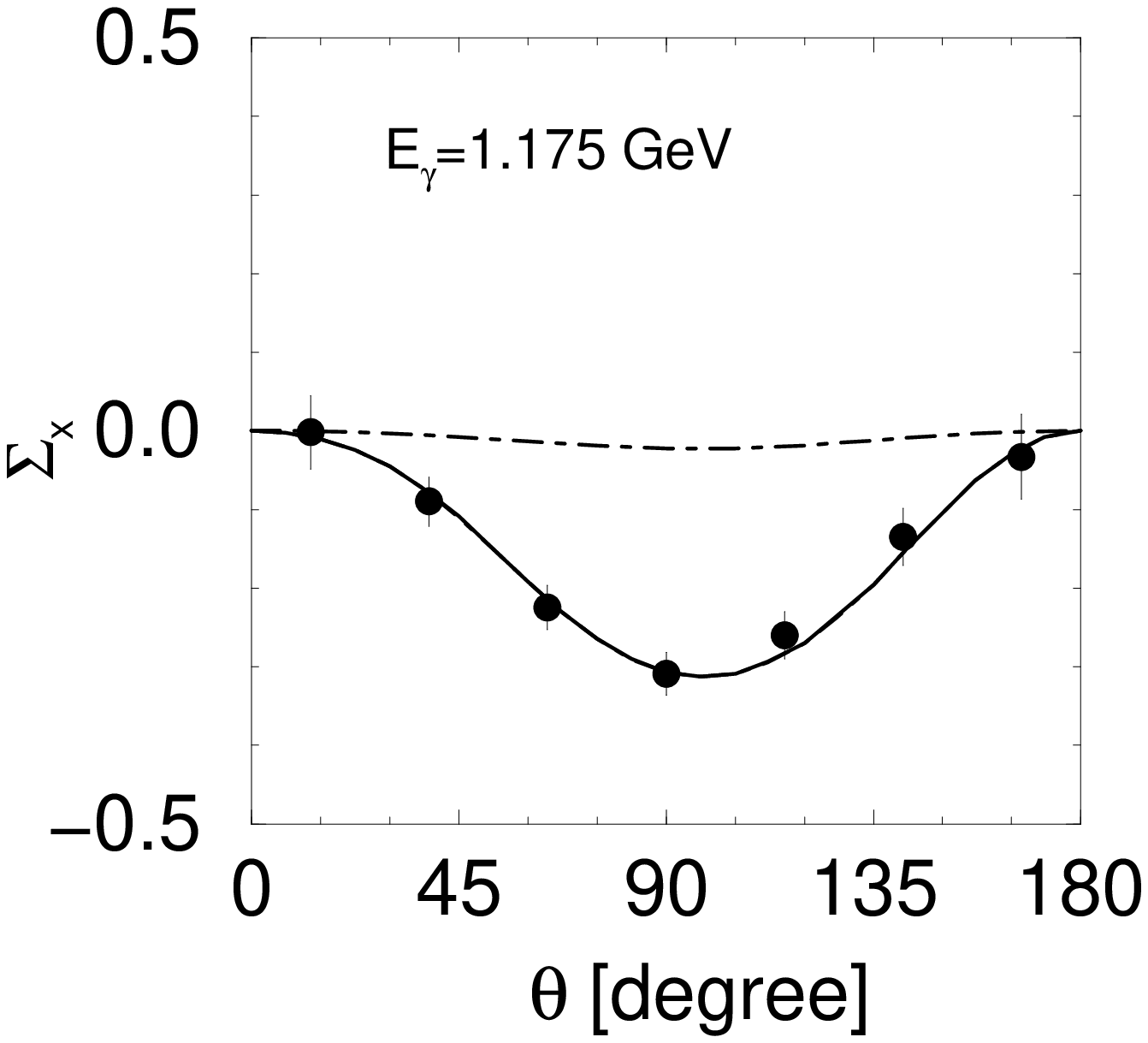, width=46mm}\qquad
 \epsfig{file=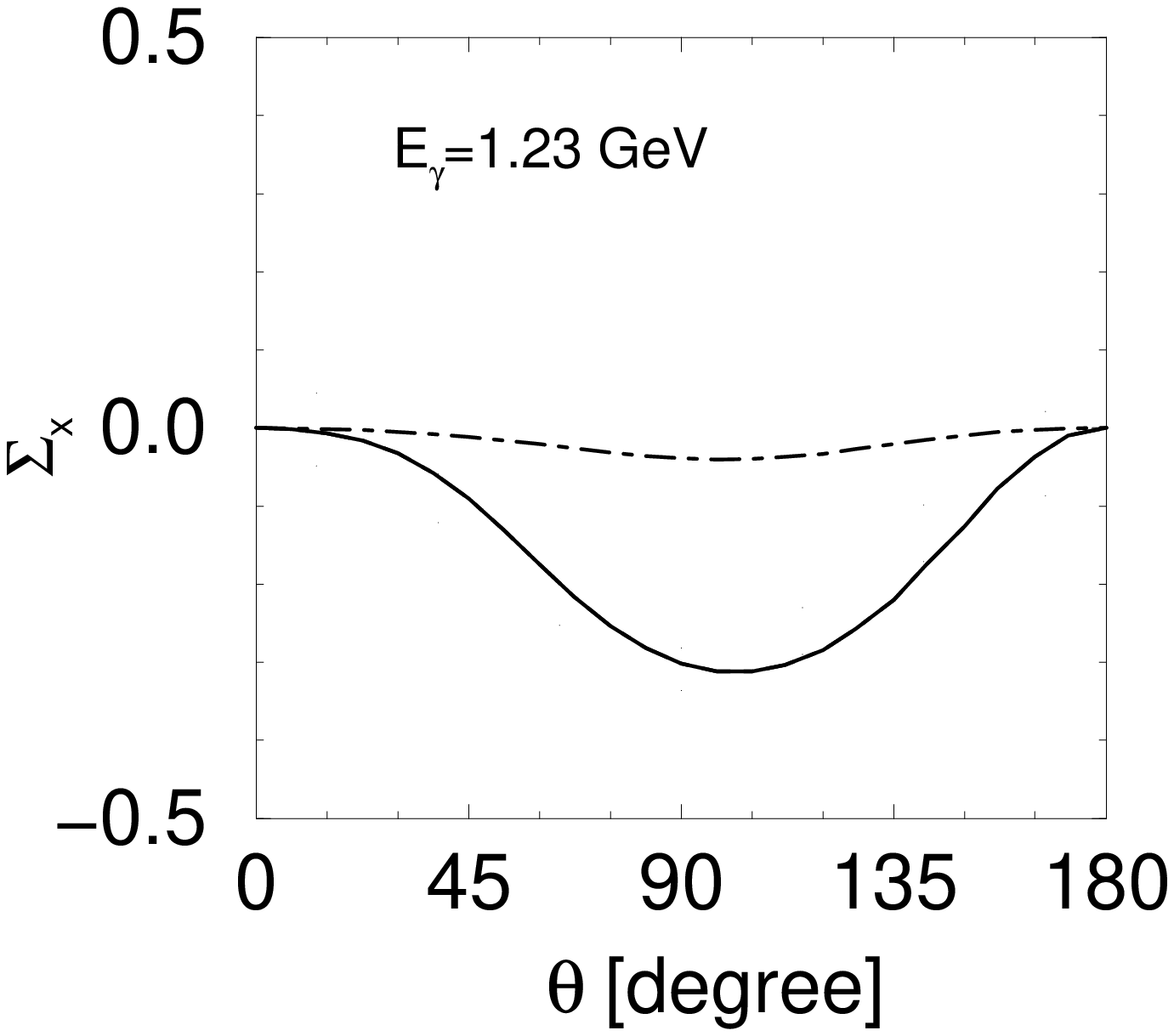, width=48mm} \vspace*{0.5cm} \caption{
 Beam asymmetry at $E_\gamma =$ $1.125$,  $1.175$, and $1.23$ GeV
 as a function of $\omega-$ production angle. The results are from
 the non-resonant background (BG) (dot-dashed), and the full
 amplitude (solid). Data are taken from Ref. \protect\cite{GRAAL}.}
 \label{fig:6}
\end{figure}

\begin{figure}
\centering \epsfig{file=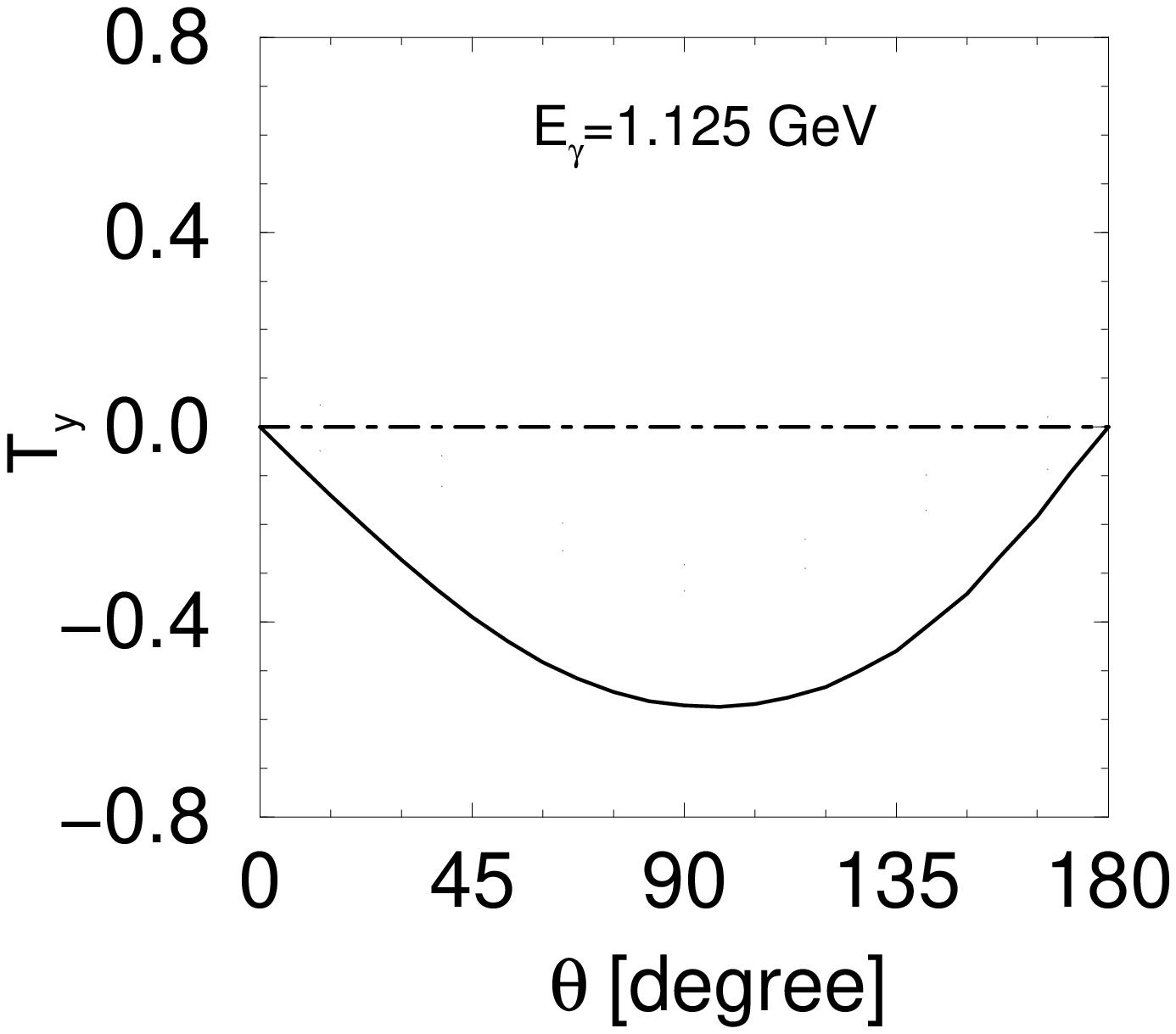, width=48mm}\qquad
\epsfig{file=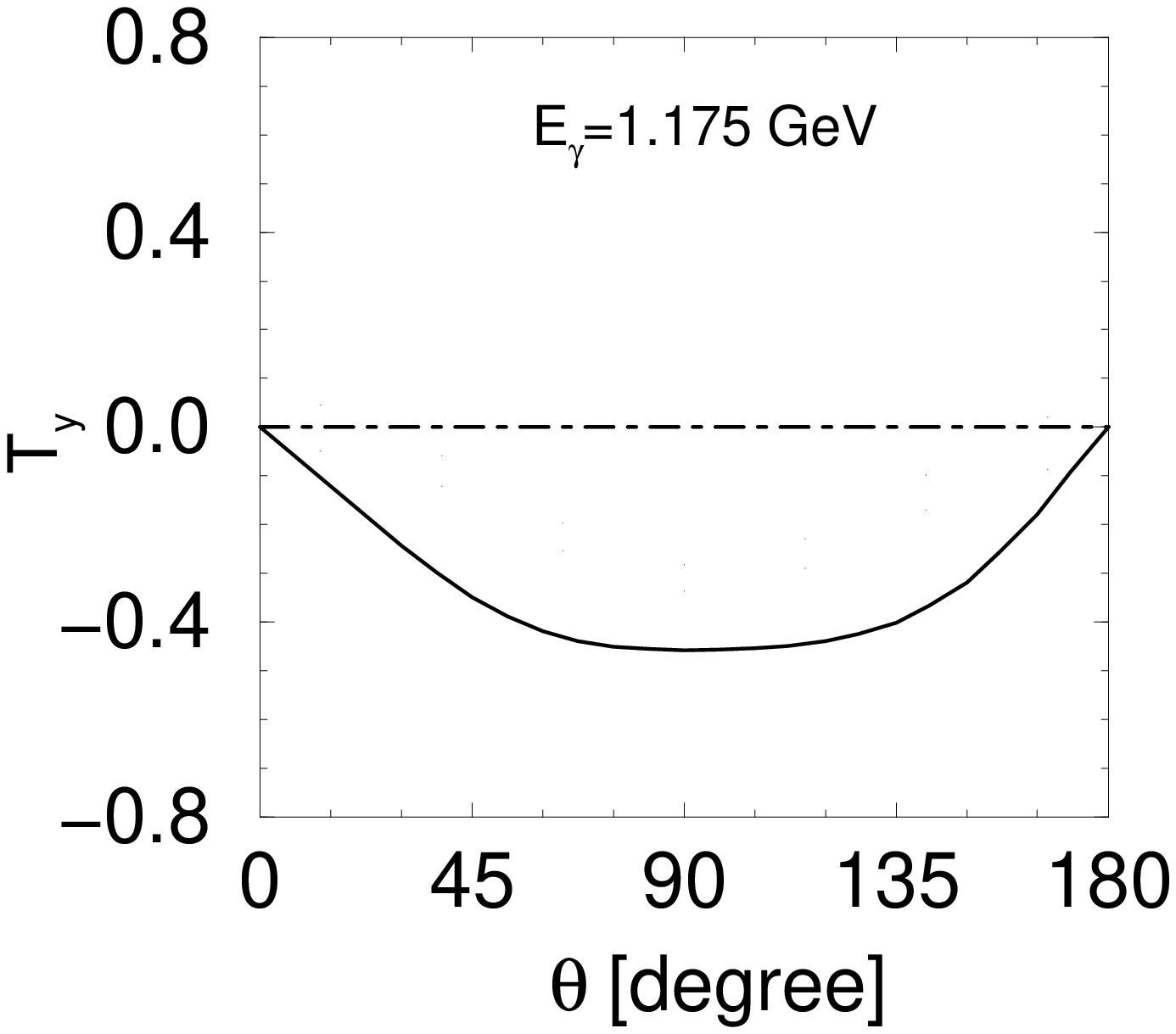, width=48mm}\qquad
\epsfig{file=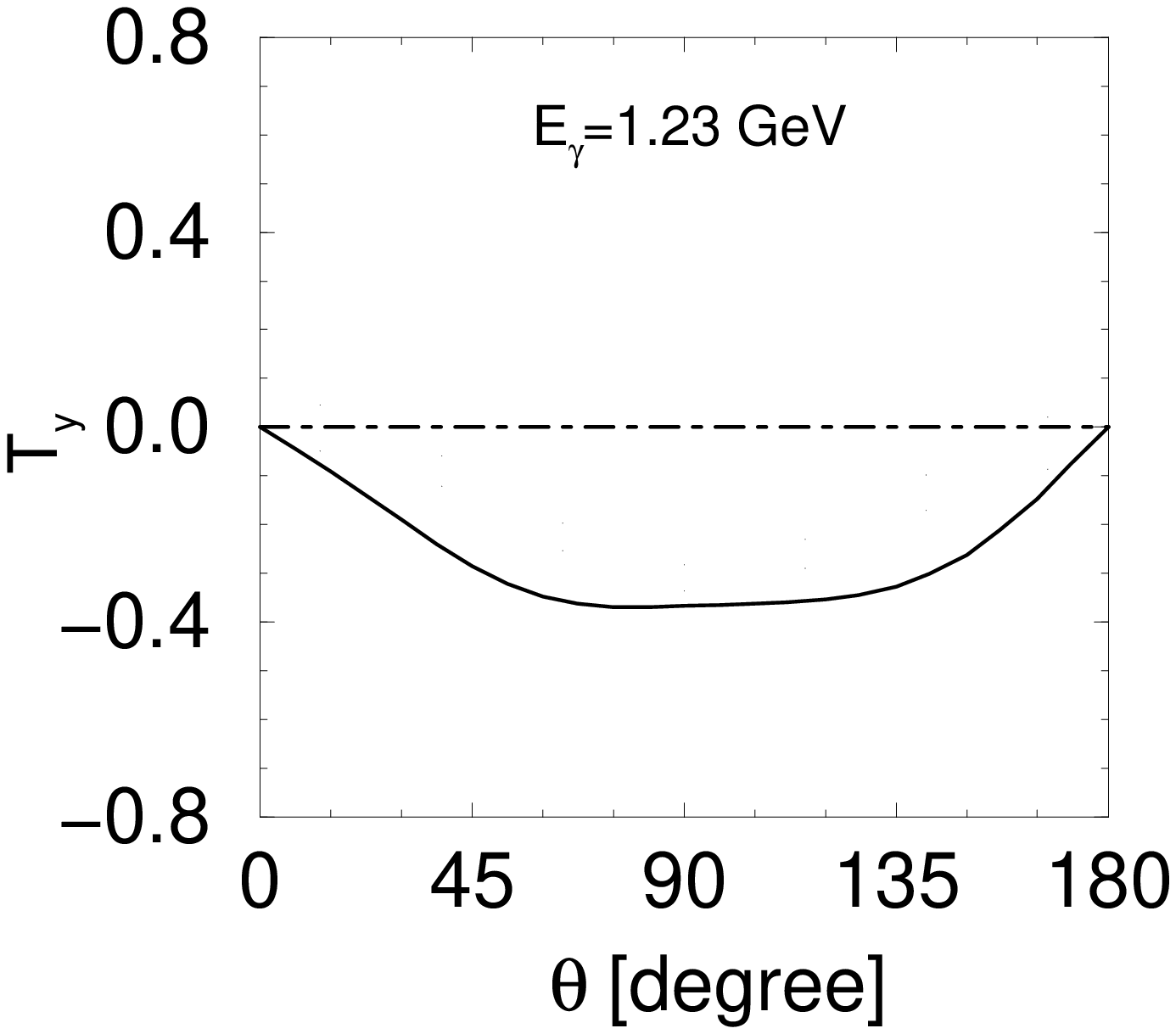, width=48mm} \vspace*{0.5cm} \caption{
Target asymmetry at $E_\gamma =$ $1.125$,  $1.175$, and $1.23$ GeV
as a function of $\omega-$ production angle. Notation is the same
as in Fig.~\protect\ref{fig:6}.} \label{fig:7}
\end{figure}

\begin{figure}
\centering \epsfig{file=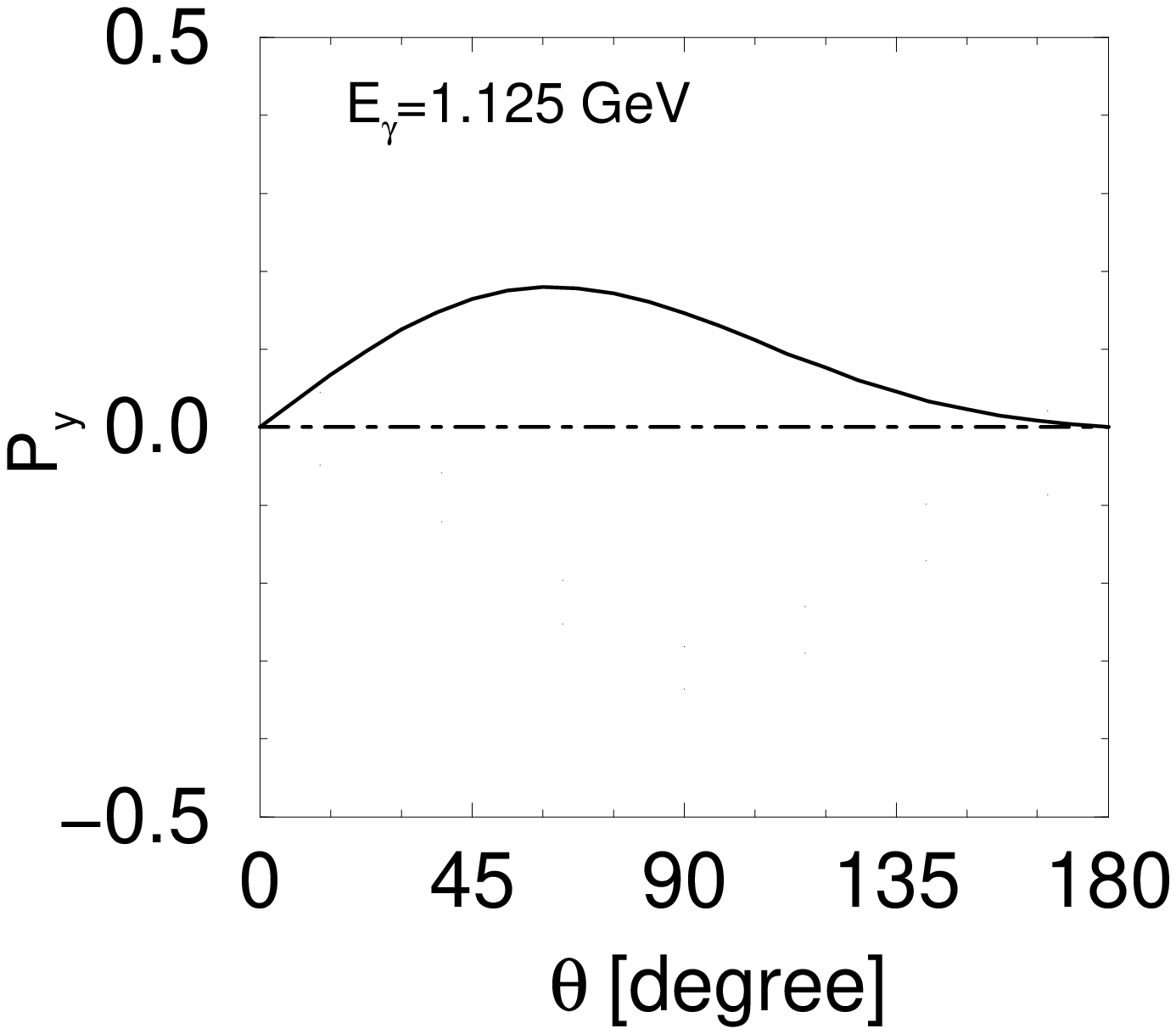, width=48mm}\qquad
\epsfig{file=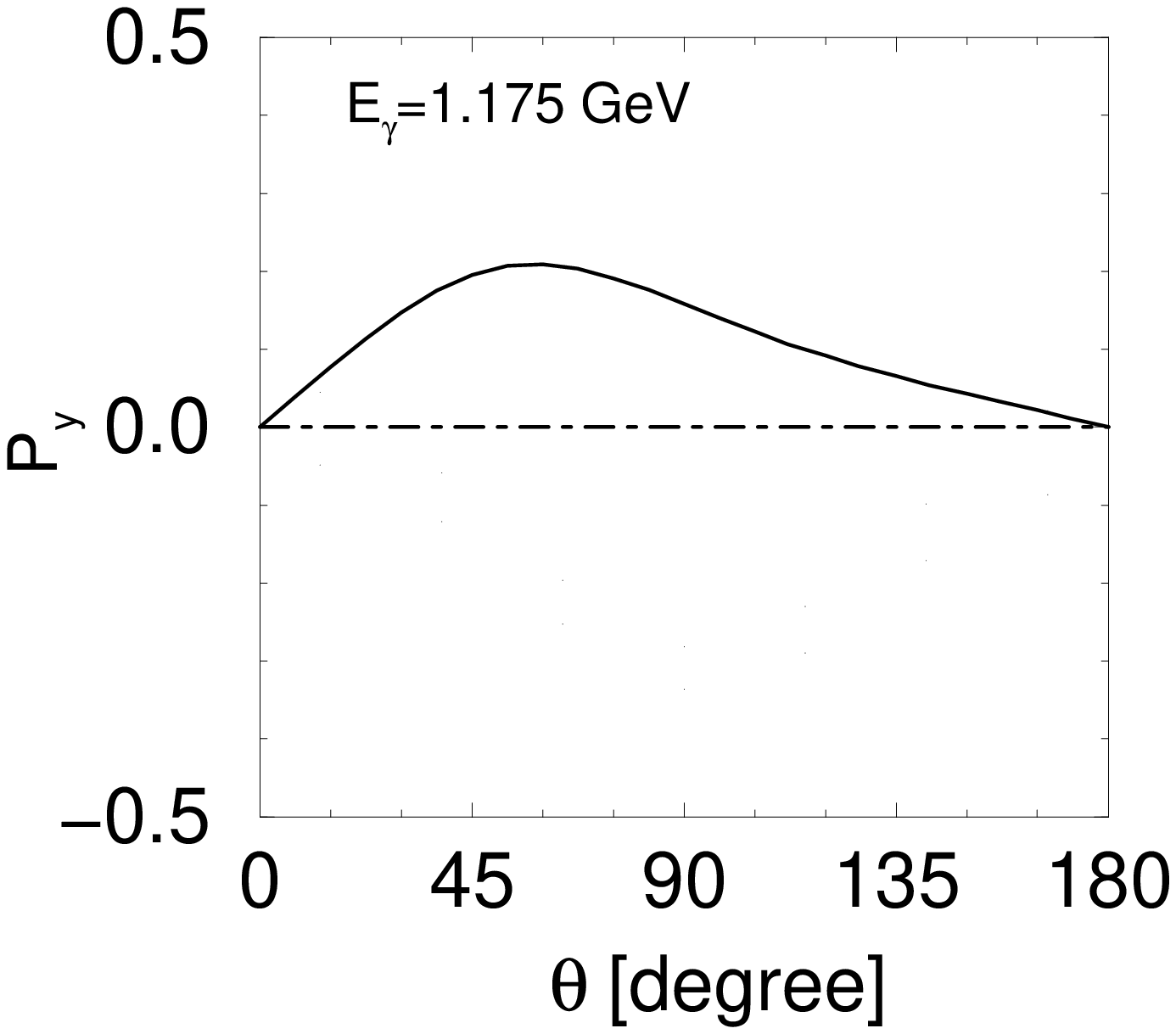, width=48mm}\qquad
\epsfig{file=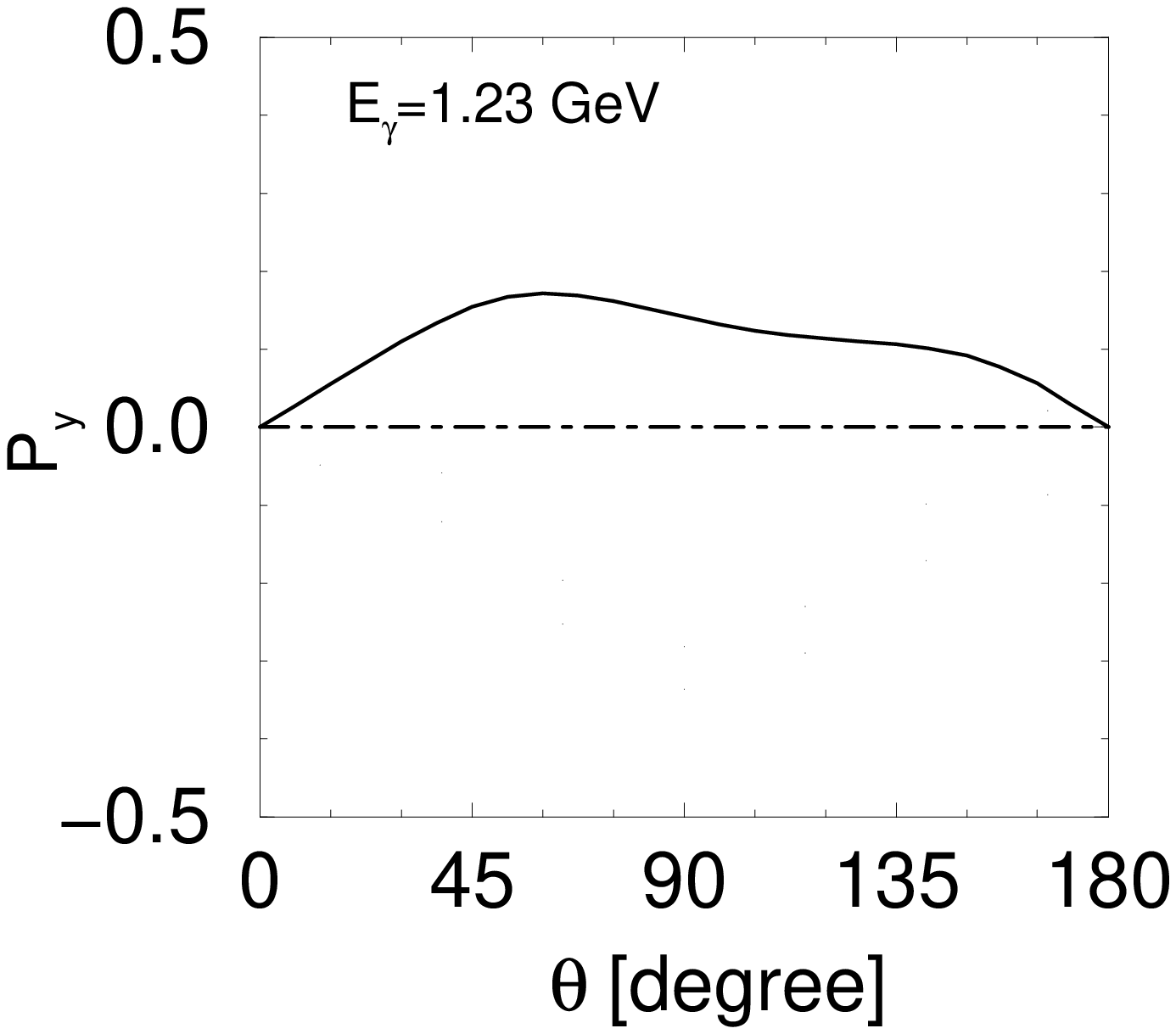, width=48mm} \vspace*{0.5cm} \caption{
Recoil asymmetry at $E_\gamma =$ $1.125$,  $1.175$, and $1.23$ GeV
as a function of $\omega-$ production angle. Notation is the same
as in Fig.~\protect\ref{fig:6}.} \label{fig:8}
\end{figure}

\begin{figure}
\centering \epsfig{file=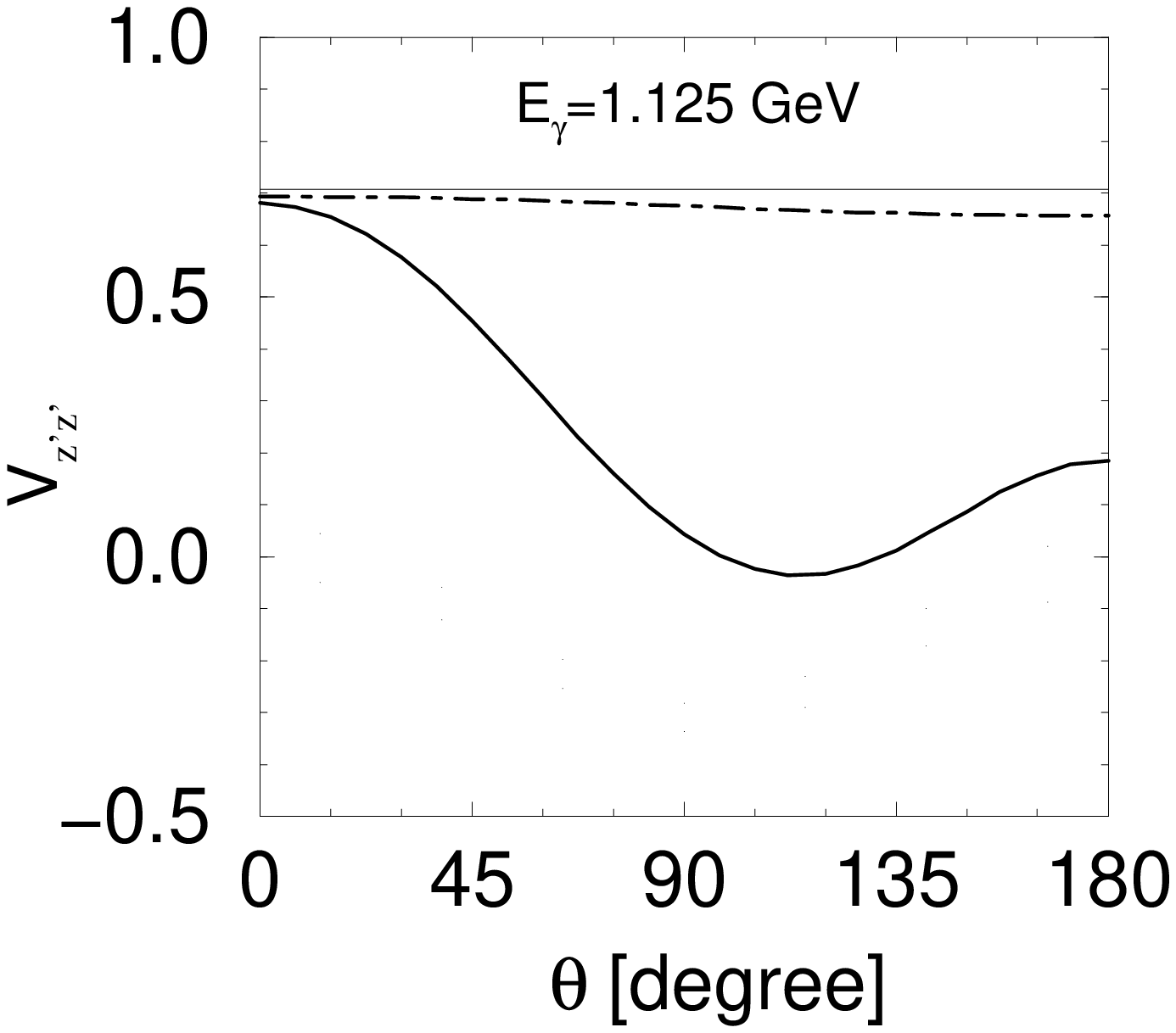, width=48mm}\qquad
\epsfig{file=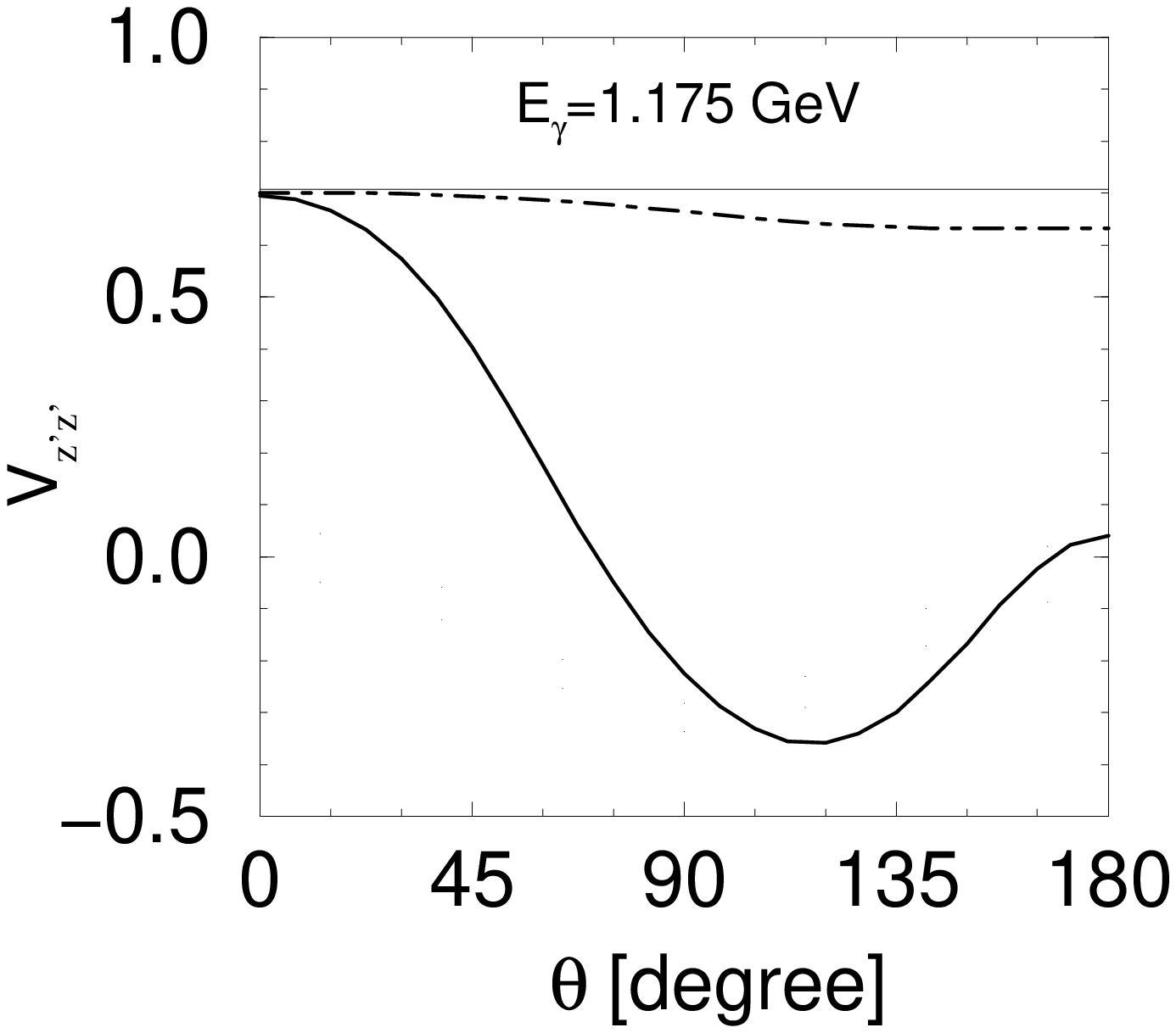, width=48mm}\qquad
\epsfig{file=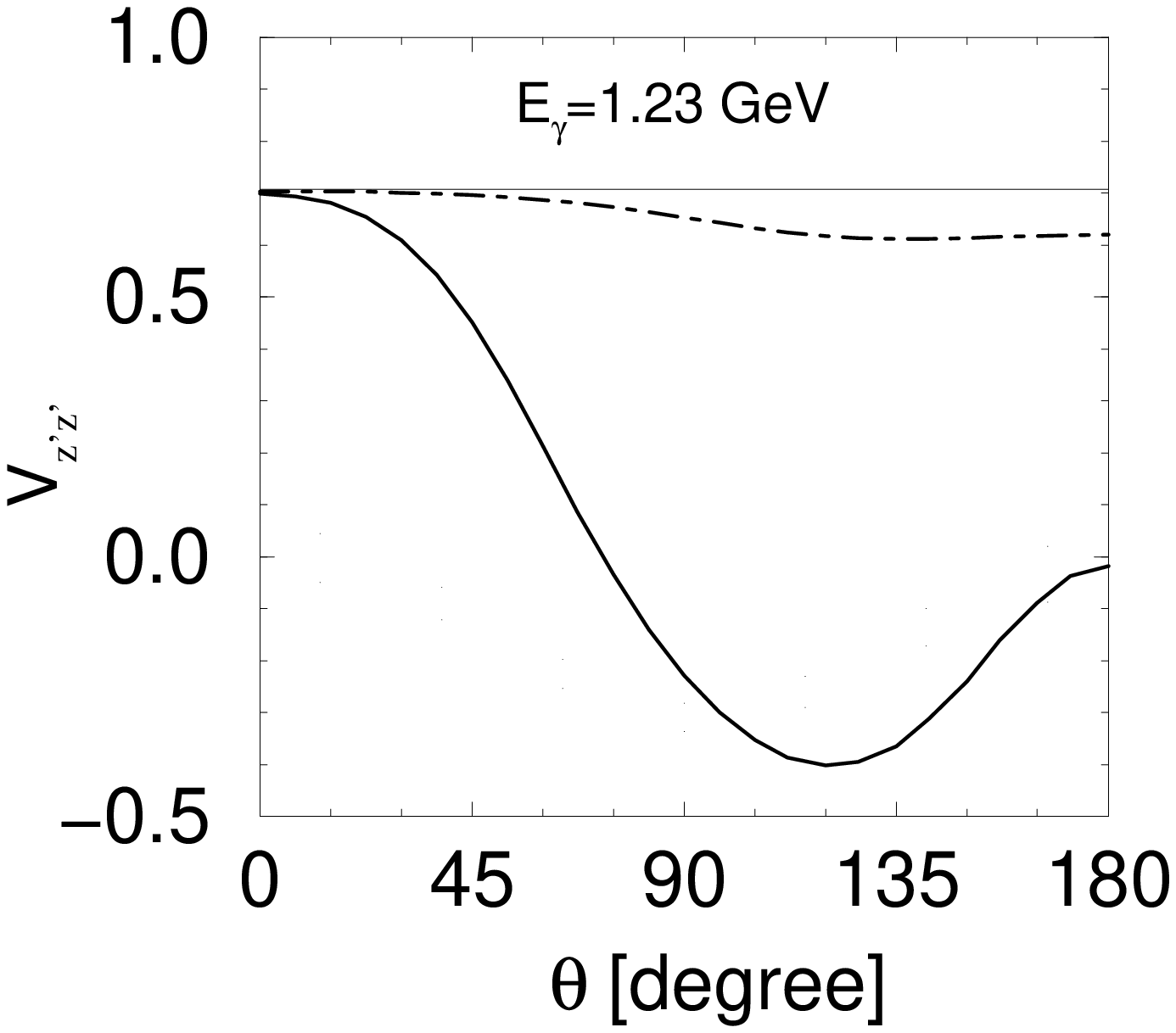, width=48mm} \vspace*{0.5cm} \caption{ The
$\omega-$meson tensor asymmetry at $E_\gamma =$ $1.125$,  $1.175$,
and $1.23$ GeV as a function of $\omega-$ production angle. The
other notation is the same as in Fig.~\protect\ref{fig:6}.}
\label{fig:9}
\end{figure}

\begin{figure}
\centering \epsfig{file=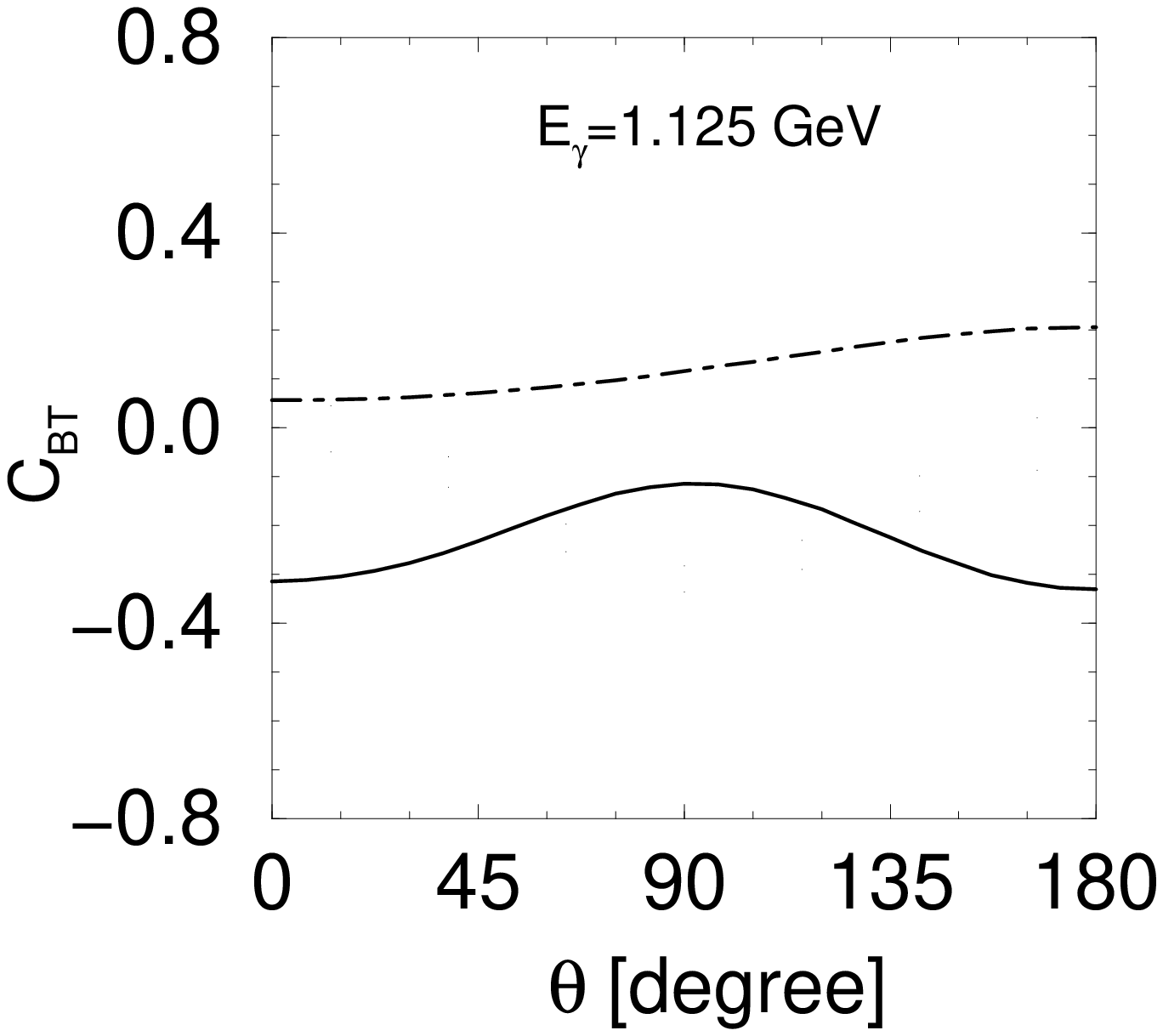, width=48mm}\qquad
\epsfig{file=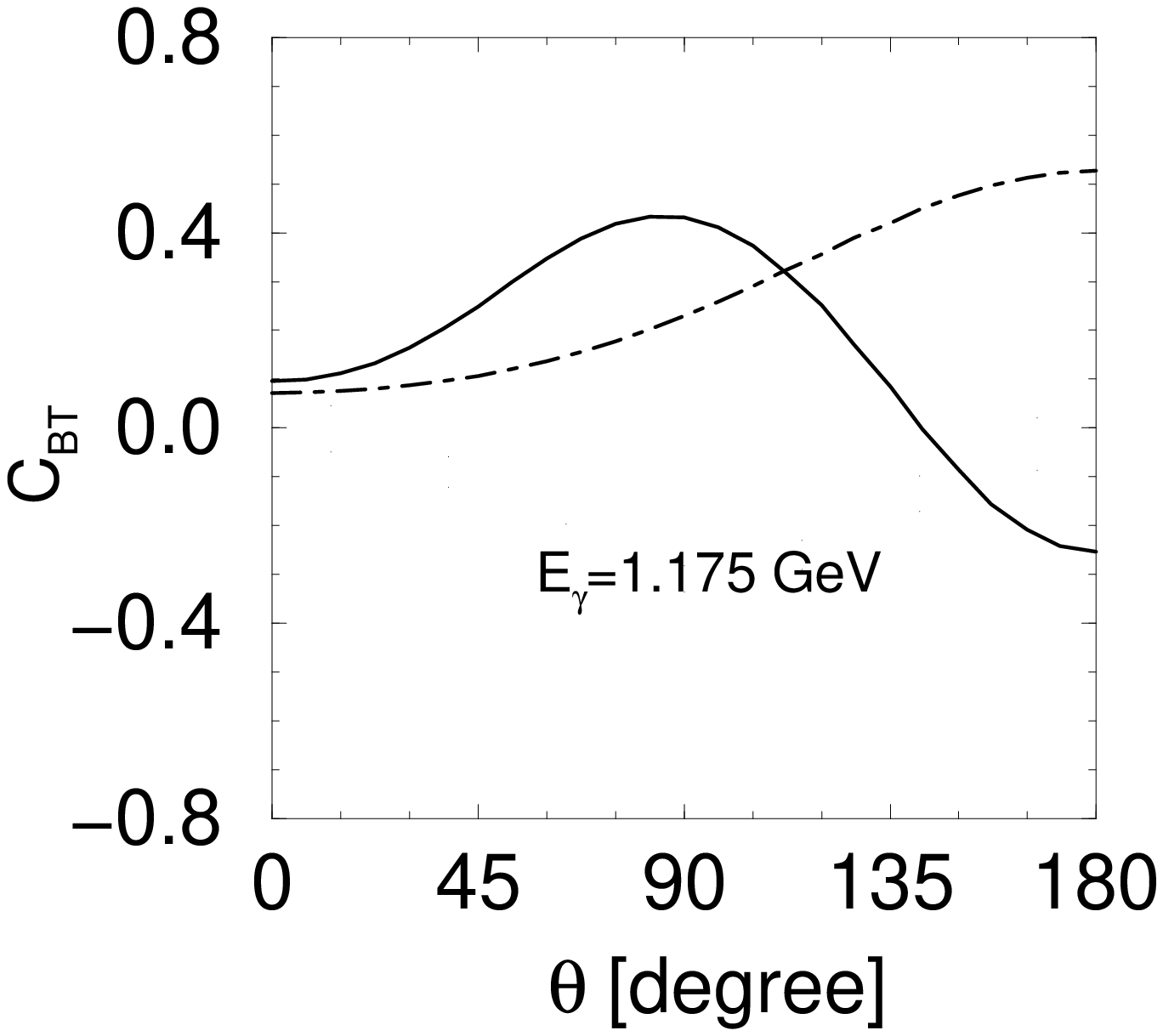, width=48mm}\qquad
\epsfig{file=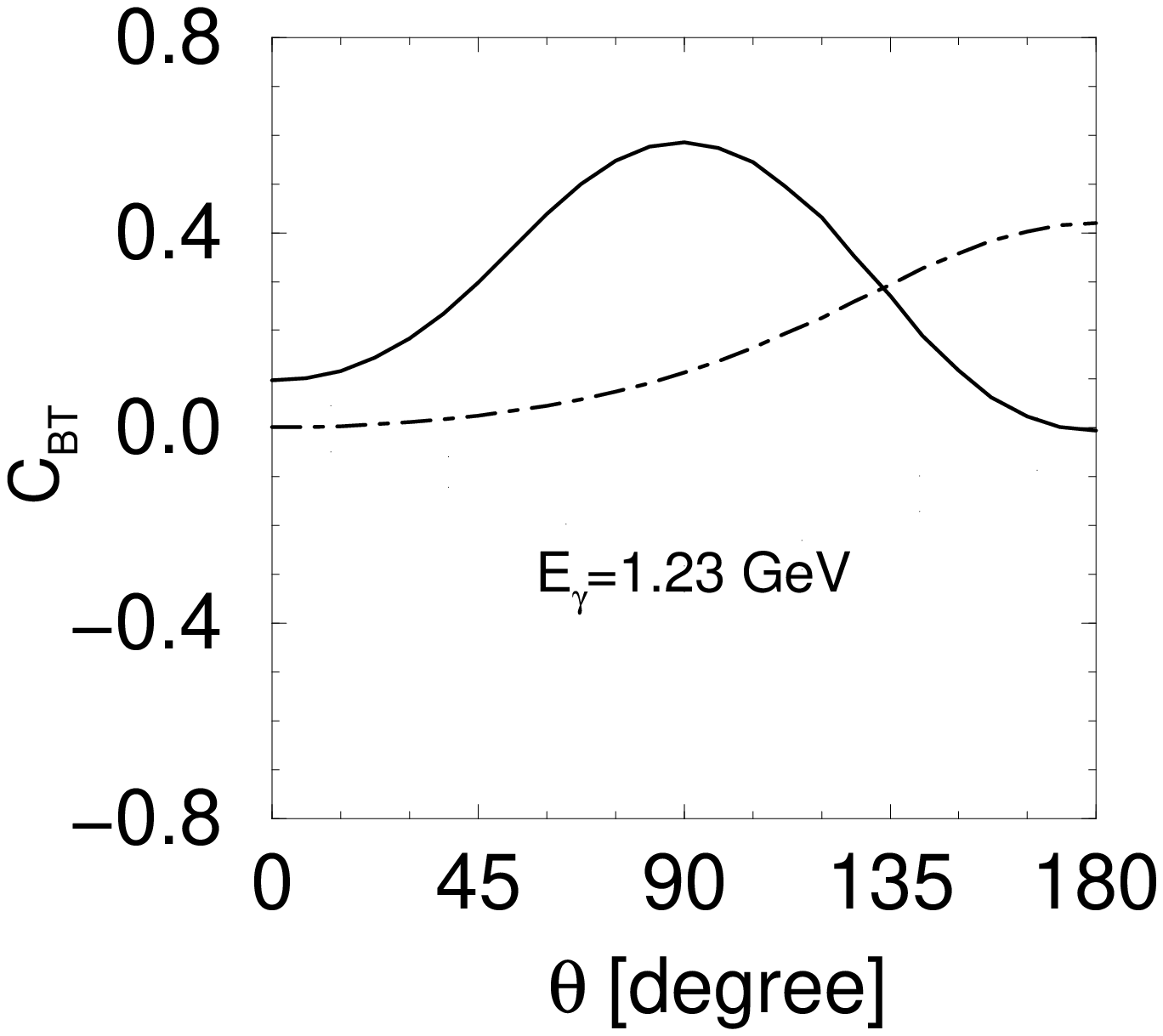, width=48mm} \vspace*{0.5cm} \caption{
Beam-target  asymmetry at $E_\gamma =$ $1.125$,  $1.175$, and
$1.23$ GeV as a function of $\omega-$ production angle. Notation
is the same as in Fig.~\protect\ref{fig:6}.} \label{fig:10}
\end{figure}

\end{document}